\documentclass[%
reprint, 
amssymb, 
aps, prb,
superscriptaddress,
]{revtex4-2}

\usepackage{newtxtext,newtxmath}

\usepackage{graphicx}
\usepackage{grffile}
\usepackage{subfigure}
\usepackage{float}

\usepackage{ulem} 
\usepackage{dcolumn}
\usepackage[super]{nth} 

\usepackage{hyperref}
\hypersetup{
    colorlinks,
    linkcolor={blue},
    citecolor={blue},
    urlcolor={blue}
}

\usepackage{physics}
\usepackage[version=3]{mhchem}

\usepackage[dvipsnames]{xcolor}

\newcommand*{\ii}{{\rm i}}
\newcommand*{\ee}{{\rm e}}

\newcommand*{\dg}{{\dagger}}

\newcommand*{\veck}{{\mathbf{k}}}

\newcommand*{\vecR}{{\mathbf{R}}}
\newcommand*{\vece}{{\mathbf{e}}}
\newcommand*{\veczero}{{\mathbf{0}}}

\newcommand*{\rmCu}{{\rm Cu}}
\newcommand*{\rmO}{{\rm O}}

\newcommand*{\AFM}{{\rm AFM}}
\newcommand*{\SC}{{\rm SC}}

\newcommand*{\mf}{{\rm mf}}
\newcommand*{\imp}{{\rm imp}}
\newcommand*{\env}{{\rm env}}
\newcommand*{\bath}{{\rm bath}}

\newcommand*{\dxy}{{$d_{x^2 - y^2}$\,}}
\newcommand*{\abinitio}{{\textit{ab initio}} }
\newcommand*{\Tc}{{$T_{\rm c}$ \,}}

\begin{document}

\preprint{Three-Band-Hub-DMET}

\title{Ground-state phase diagram of the three-band Hubbard model from density matrix embedding theory}

\author{Zhi-Hao Cui}
\affiliation{Division of Chemistry and Chemical Engineering, California Institute of Technology, Pasadena, California 91125, United States}

\author{Chong Sun}
\affiliation{Division of Chemistry and Chemical Engineering, California Institute of Technology, Pasadena, California 91125, United States}

\author{Ushnish Ray}
\affiliation{Division of Chemistry and Chemical Engineering, California Institute of Technology, Pasadena, California 91125, United States}

\author{Bo-Xiao Zheng}
\affiliation{AxiomQuant Investment Management LLC, Shanghai 200120, China}
\affiliation{Division of Chemistry and Chemical Engineering, California Institute of Technology, Pasadena, California 91125, United States}
\affiliation{Department of Chemistry, Princeton University, Princeton, New Jersey 08544, United States}

\author{Qiming Sun}
\affiliation{AxiomQuant Investment Management LLC, Shanghai 200120, China}
\affiliation{Division of Chemistry and Chemical Engineering, California Institute of Technology, Pasadena, California 91125, United States}

\author{Garnet Kin-Lic Chan}
 \email{gkc1000@gmail.com}
\affiliation{Division of Chemistry and Chemical Engineering, California Institute of Technology, Pasadena, California 91125, United States}

\date{\today}

\begin{abstract}
We determine the ground-state phase diagram of the three-band Hubbard model 
across a range of model parameters using density matrix embedding theory.
We study the atomic-scale nature of the antiferromagnetic (AFM) and
superconducting (SC) orders, explicitly including the oxygen degrees of freedom. 
All parametrizations of the model display AFM and SC phases, but
the decay of AFM order with doping is too slow compared to the experimental phase diagram,
and further, coexistence of AFM and SC orders occurs in all parameter sets.
The local magnetic moment localizes entirely at the copper sites.
The magnetic phase diagram is particularly sensitive to $\Delta_{pd}$ and $t_{pp}$, and 
existing estimates of the charge transfer gap $\Delta_{pd}$ appear too large in so-called minimal model parametrizations.
The electron-doped side of the phase diagram is qualitatively distinct from the hole-doped side and
we find an unusual two-peak structure in the SC in the full model parametrization.
Examining the SC order at the atomic scale, within the larger scale $d_{x^2 - y^2}$-wave SC pairing order between Cu-Cu and O-O,
we also observe a local $p_{x (y)}$ [or $d_{xz (yz)}$]-symmetry modulation of the pair density on the Cu-O bonds.
Our work highlights some of the features that arise in a three-band versus one-band picture,
the role of the oxygen degrees of freedom in new kinds of atomic-scale SC orders,
and the necessity of re-evaluating current parametrizations of the three-band Hubbard model.
\end{abstract}


\maketitle


\section{Introduction}\label{sec: introduction}


The three-band Hubbard model, also known as the Emery model~\cite{Emery87}, is
generally believed to contain the essential physics of the high-\Tc cuprates
 that arises from the interplay between the copper \dxy and oxygen $p_{x(y)}$
 orbitals in the \ce{CuO2} layers.
 Given the complexity of the model, 
 commonly, 
 the three-band model is further simplified and several 
simpler low-energy Hamiltonians have been proposed, such as the
one-band Hubbard model~\cite{Hubbard63, Anderson87}, $t$-$J$
model~\cite{Anderson87, Zhang88ZRS}, and two-band model~\cite{Sakakibara10}. The
first two are effective one-band models and are equivalent in the
strong-coupling limit. In particular, the two-dimensional (2D) one-band Hubbard model has been
extensively investigated using various numerical approaches (see Refs. ~\cite{LeBlanc15,
Zheng17sci} and the references therein). Much of the physics seen in 
high-\Tc materials, e.g. $d$-wave pairing, density waves, the pseudogap phase
and stripe order, has been observed in studies of the simpler one-band Hubbard model
within certain ranges of parameters ~\cite{Zheng17sci}. 

However, despite the progress in understanding the one-band Hubbard model and its variants, there are still
important reasons to 
go beyond the  one-band picture to study the original three-band model directly.
For instance, (a) some important physics may be lost in the reduction to the one-band approximation
(such as a role for the oxygen degrees of freedom in the pseudogap phase~\cite{Fauque06}), (b) 
near degeneracies of competing states seen in the one-band case~\cite{Zheng17sci} may in fact be resolved
with the additional degrees of freedom of the three-band model, and (c) the 
three-band model retains the atomic structure of  the \ce{CuO2} layer and thus has a direct link to the
structure of real materials as well as experimental measurements of orders at the atomic scale.
Previously, the three-band Hubbard model has been investigated with several numerical
methods, including direct simulations of finite lattices [by exact
diagonalization (ED) ~\cite{Hybertsen90, Scalettar91, Cini97,
Greiter07-3band-ED-orb-current, Thomale08-3band-ED-orb-current, Shirakawa13,
Kung14-3band-ED-QMC-orb-current}, quantum Monte Carlo (QMC)~\cite{Dopf90, Kuroki96, Guerrero98,
Yanagisawa01, Weber14-3band-QMC, Kung14-3band-ED-QMC-orb-current, Kung16, Huang17, Vitali19-3band}, 
density matrix renormalization group (DMRG) ~\cite{Jeckelmann98, Nishimoto02,
White15, Huang17}, and the random phase
approximation~\cite{Bulut13, Maier14, Atkinson15}] and via Green's function based embedding theories [such as dynamical mean-field
theory (DMFT) and its cluster extensions~\cite{Maier99, Zolfl00, Kent08, Medici09, Weber12, Go15}, and the variational cluster approximation
(VCA)~\cite{Arrigoni09, Hanke10}]. 
However, due to the complexity of the model, unlike in the one-band case, a consensus on much of the physics has yet to be reached.


Over the past few years, density matrix embedding theory (DMET) ~\cite{Knizia12} 
has emerged as a powerful cluster embedding method. 
The basic idea of DMET is similar to that of (cellular) DMFT in the sense that they 
both map an infinite lattice to an impurity model with an environment that can be described by 
bath degrees of freedom, and the impurity model is self-consistently improved by matching physical quantities 
between a single-particle lattice solution and the correlated cluster (impurity) calculation. 
Technically, however, DMET has a different structure to Green's function based embedding methods,
and is formulated without frequency dependence and with a finite
set of  bath orbitals (bounded by the number of impurity orbitals). The lack of frequency-dependent quantities 
means that DMET calculations can utilize efficient ground-state impurity solvers that can typically treat larger clusters
than can be addressed by solvers that target the impurity Green's function.
DMET has been applied to a wide range of fermionic lattice models
~\cite{Knizia12, Bulik14, Zheng16, Zheng17, Zheng17sci, Wu19pdmet}, 
\abinitio chemical Hamiltonians~\cite{Knizia13, Wouters16, Pham18,
Bulik14detsolid, Cui20-dmet-solid, Pham20-dmet-solid}, 
and nonfermionic systems~\cite{Fan15, Sandhoefer16}, 
as well as excited states~\cite{Booth15, Sun20ftdmet} and time-dependent problems~\cite{Kretchmer18}. 
For a detailed review of DMET, we refer to Ref. ~\cite{Wouters16}. 

In earlier work, DMET successfully provided an accurate description of the ground-state orders of the one-band Hubbard model~\cite{Zheng16}, 
including in the difficult underdoped region~\cite{Zheng17sci}. In this work, we therefore
attempt to understand the more complicated three-band Hubbard model using
DMET. As we shall see, we can use DMET to provide a detailed description 
of the ground-state phases and orders as a function of doping, including the doping asymmetry and atomic-scale orders
that are new to the three-band case. Another complication of the three-band model is
 the much larger parameter space than the one-band case. We use both existing parametrizations that
have been published in the literature, as well as explicitly model the influence of different individual parameters
on the orders. Our findings provide insights
into the detailed picture of magnetic and superconducting orders that is provided by three-band models.


\section{Models and Methods}\label{sec: theory}

\subsection{Model parametrization}

\begin{figure}[!htb]
\subfigure{\label{fig: cluster}\includegraphics[width=40mm,  clip]{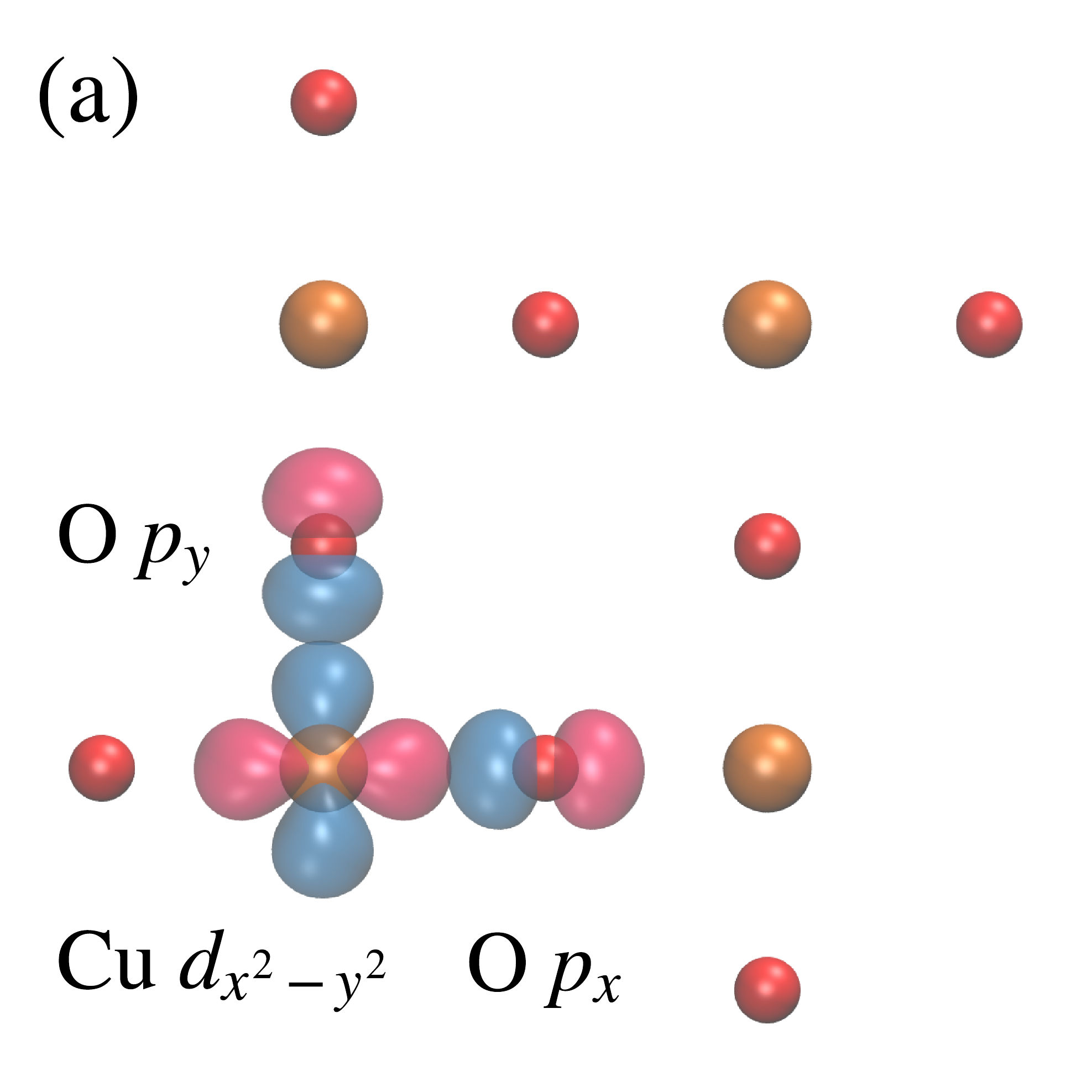}}
\subfigure{\label{fig: model param}\includegraphics[width=40mm,  clip]{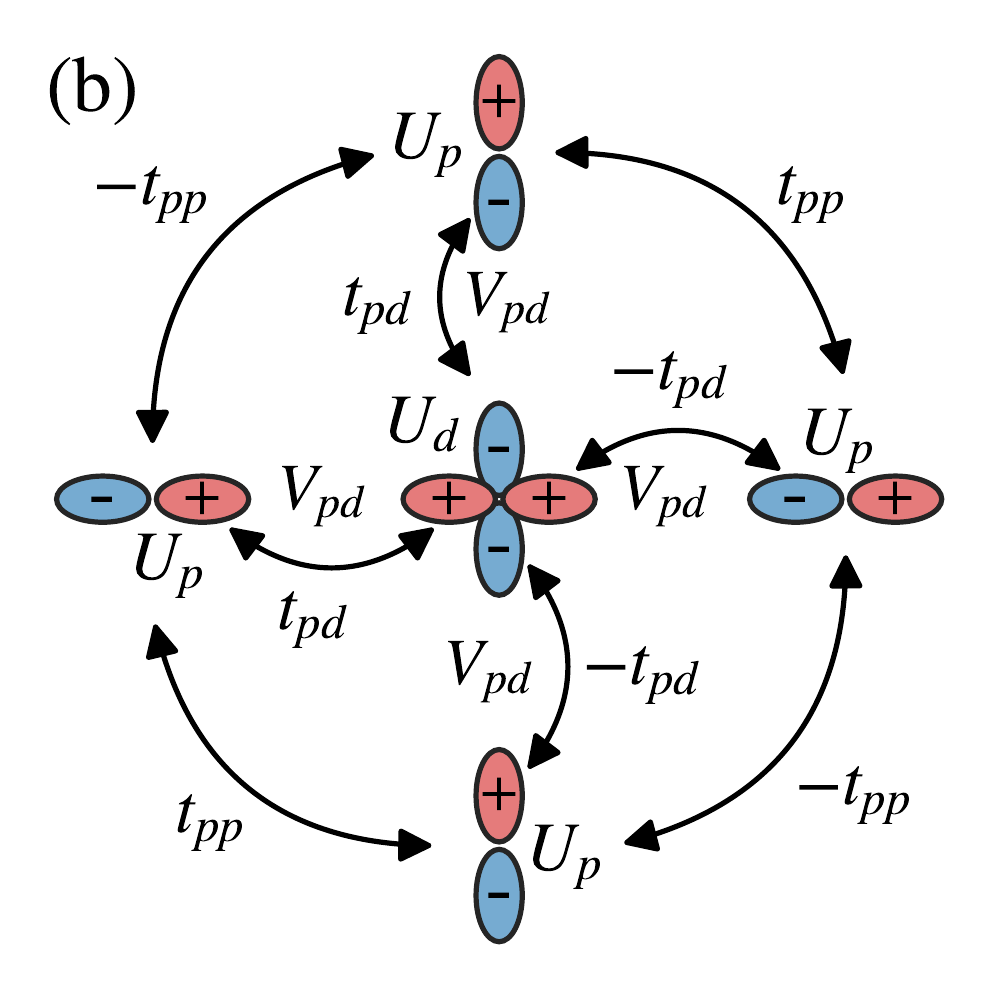}}
\caption{An illustration of the three-band Hubbard model: 
(a) the symmetric cluster used in the DMET calculations, 
where the orange and red atoms denote copper and oxygen respectively; 
(b) definition of the model parameters and the phase convention. }
\end{figure}

As a minimal atomic model of the \ce{CuO2} layer in cuprates, 
the three-band model describes the on-site and nearest-neighbor interactions among the
Cu $d_{x^2-y^2}$ and O $p_{x}$, $p_{y}$ orbitals [see Fig. \ref{fig: cluster}].
In the hole representation, the Hamiltonian reads,
\begin{equation}\label{eq: three band model}
\begin{split}
\mathcal{H} &= t_{pd} \sum_{\expval{ij} \sigma} \qty(d^{\dagger}_{i\sigma} p_{j\sigma} + \mathrm{H.c.}) + t_{pp} \sum_{\expval{jj'} \sigma} \qty(p^{\dagger}_{j\sigma} p_{j'\sigma} + \mathrm{H.c.}) \\
&-\Delta_{pd} \sum_{i \sigma} n^d_{i \sigma} + U_{d} \sum_{i} n^d_{i\alpha} n^d_{i\beta}  +  U_{p} \sum_{j} n^p_{j\alpha} n^p_{j\beta} \\
&+ V_{pd} \sum_{\expval{ij} \sigma \sigma'} n^{d}_{i \sigma} n^{p}_{j \sigma'},  \\
\end{split}
\end{equation}
where $\expval{\cdots}$ denotes nearest neighbors, 
$d_{i\sigma}^{(\dagger)}$ and $p_{j\sigma}^{(\dagger)}$ destroy (create) a hole 
with spin $\sigma$ ($\in \{\alpha, \beta\}$) on the Cu $d$ and O $p$ orbitals respectively, 
$n^{d}_{i\sigma}$ and $n^{p}_{j\sigma}$ are the corresponding hole particle-number operators, 
and the charge transfer gap $\Delta_{pd}$ is defined as the orbital energy difference, 
$\epsilon_{p} - \epsilon_{d}$.  Similarly to in the one-band Hubbard model, 
the hopping term and on-site Coulomb repulsion will be denoted  $t$ and $U$, 
and the Coulomb interaction between  nearest-neighbor $p$, $d$ orbitals will be denoted $V_{pd}$. 
Note that the hopping term involves a phase factor ($\pm 1$) introduced by the choice of orbital 
orientation in the basis as shown in Fig. \ref{fig: model param}. 

There has been much work to determine the parameters of the three-band model; however, a
consensus set does not exist ~\cite{Hybertsen89, McMahan90, Martin963band,
Kent08, Hirayama18}. There has been particular debate about the size of the
charge transfer gap $\Delta_{pd}$ ~\cite{Kent08, Chiciak18}. 

\begin{table}[!htb]
\caption{\label{tab:parameter}%
Parameters of the three-band Hubbard model used in this work, in units of eV. 
The parameters correspond to the hole representation. 
    }
\begin{ruledtabular}
\begin{tabular}{lcccccc}
Model & $t_{pd}$ & $\Delta_{pd}$ & $U_{d}$ & $t_{pp}$ & $U_p$ & $V_{pd}$   \\
\hline
Hybertsen\footnotemark[1]  & 1.3	& 3.6  &	10.5 &	 	&	   	&      \\
Martin\footnotemark[2]     & 1.8	& 5.4  &	16.5 &	 	&	   	&      \\
Hanke\footnotemark[3]      & 1.5  	& 4.5  &	12.0  &	 	&	   	&      \\
Hanke full\footnotemark[3] & 1.5	& 4.5  &	12.0  &  0.75	&	5.25 &  0.75  \\
\end{tabular}
\end{ruledtabular}
\footnotetext[1]{From Ref.~\cite{Hybertsen89}.}
\footnotetext[2]{From Ref.~\cite{Martin963band}.}
\footnotetext[3]{From Ref.~\cite{Hanke10}.}
\end{table}
In this work, we consider four sets of published model parameters, see Table \ref{tab:parameter}, as well
as the sensitivity of orders to changing these parameters.
Note that all parameter sets are given in eV, thus all energies in this work are reported in units of eV unless otherwise specified. 
The first three sets include only the most essential terms, i.e.,  $t_{pd}$, $U_{d}$ and $\Delta_{pd}$, 
and thus we refer to them as \textit{minimal} parametrizations. 
When normalized to units of $t_{pd}$, the other parameters vary within a range of 10\%. 
The fourth set involves all terms in Eq. \eqref{eq: three band model}. 
We refer to this as a \textit{full} parametrization. In the hole representation, the minimal parametrization is equivalent to the full model with $t_{pp}$, $U_{p}$ and $V_{pd}$ set to zero.

\subsection{Computational formulation}\label{subsec: computational details}

\subsubsection{Framework}
DMET approximates the expectation values in the interacting lattice by those in a quantum impurity model.
The impurity model is solved simultaneously with a fictitious non-interacting lattice problem, whose
ground state defines the bath sites of the impurity model via a Schmidt
decomposition ~\cite{Peschel12, Knizia12}.
Self-consistency is achieved by matching the one-particle density matrix of the impurity model 
and non-interacting lattice ground states via a \textit{correlation potential} $u$ applied to the non-interacting lattice. 
The basic steps of the DMET self-consistency loop are thus
(a) compute the ground state of the non-interacting lattice Hamiltonian with correlation potential $u$, 
(b) construct the bath sites and  impurity model Hamiltonian, 
(c) solve for the ground state of the impurity model, and 
(d) match the one-particle density matrices of the lattice Hamiltonian and impurity model to update $u$. 
The cycle ends when the correlation potential $u$ is converged. 

In this work, we are interested in both magnetic and superconducting phases. 
Consequently, the correlation potential takes the form 
\begin{equation}\label{eq: u potential}
u = \sum_{ij\sigma} v^{\sigma}_{ij} a^{\dg}_{i\sigma} a_{j\sigma} + \sum_{ij} \Delta^{\alpha\beta}_{ij} a^{\dg}_{i\alpha} a^{\dg}_{j\beta} + \mathrm{H.c.} ,
\end{equation}
where optimizing over $v^{\sigma}$ and $\Delta^{\alpha\beta}$ in the
self-consistency procedure allows for formation of spin polarized and
singlet superconducting pairing (between two spin channels $\alpha$ and $\beta$) order in the lattice and impurity problems. 
The non-interacting lattice Hamiltonian is then of Bogoliubov-de Gennes form
~\cite{deGennes66}.
The corresponding ground-state solution is a mean-field
Bardeen-Cooper-Schrieffer (BCS) wavefunction, and a set of bath orbitals 
that describes the environment can be constructed from the corresponding generalized density
matrix. The detailed formulas for the bath construction are summarized in Appendix \ref{app : DMET bath construction} and the integral transformation
expressions for the BCS mean field can be found
in Refs. ~\cite{Zheng16, Zheng18}. These routines have been implemented in \textsc{libDMET} \cite{libdmet, libdmetsolid}.


\subsubsection{Impurity and lattice}
We used a $2 \times 2$ impurity cluster of CuO$_2$ primitive cells~\cite{Hanke10} which retains the inversion and
four-fold rotation symmetry of the lattice [see Fig \ref{fig: cluster}]. We embedded the cluster in a $20 \times 20$
unit-cell ($40 \times 40$ site-length) lattice. 
We performed DMET calculations for dopings $x$ ranging between -0.8 and 0.8 (negative denotes electron doping and positive denotes hole doping). Unless otherwise specified, we initialized $u$ with an antiferromagnetic guess and a random pairing potential.

\subsubsection{Impurity Hamiltonian and solver}
The impurity model Hamiltonian was constructed
using the non-interacting DMET bath formalism~\cite{Knizia12, Zheng16}, 
and the ground state was determined using a density matrix renormalization
group (DMRG) solver ~\cite{White92, White93}, 
allowing for particle number symmetry breaking and spin polarization
~\cite{Zheng16}. 
During the DMET self-consistent cycle we used a maximum bond dimension $M = 800$. 
Subsequent bond dimension convergence checks were performed using (up to) $M = 2000$. 
To minimize entanglement and ensure a small bond dimension $M$ in the ground state, 
we rotated the impurity Hamiltonian into a basis of split-localized molecular orbitals (MOs) obtained from the self-consistent
Hartree-Fock-Bogoliubov (HFB) method, 
where the occupied and virtual MOs were computed using the \textsc{PySCF}
package ~\cite{Sun18pyscf, Sun20pyscf},
and the occupied and virtual spaces were subsequently localized separately using
the Edmiston-Ruedenberg procedure that maximizes the Coulomb energy of each
orbital ~\cite{Edmiston63, Chan02}. 
The standard genetic algorithm implemented in the~\textsc{Block} program~\cite{Chan02, Chan04, Chan11, Sharma12} 
was used to order the orbitals for the DMRG calculation. 
The tolerance of the DMRG sweep energy was set to $10^{-6}$. Convergence
checks on the accuracy of DMRG energies are described in Appendix \ref{app: convergence}.

\subsubsection{DMET self-consistency}
We carried out DMET self-consistency using full
impurity-bath fitting ~\cite{Zheng16, Wouters16}, 
where the cost function measures the least-squares difference between the correlated one-particle density matrix $\gamma^{\rm corr}$ and
the non-interacting lattice density matrix projected to the full impurity problem $\gamma^{\rm mf}$,
\begin{equation}\label{eq: cost function}
w (u) = \sum^{\imp + \bath}_{kl} \qty[\gamma^{\mf}_{kl} (u) - \gamma^{\rm corr}_{kl}]^2.
\end{equation}
We minimized $w$ using a conjugate gradient (CG) minimizer with line search. 
Since the gap of the non-interacting lattice model is often small (in the case of doped systems), a finite inverse temperature $\beta = 1000$ $t_{pd}$ was used to define the  non-interacting density matrix 
to ensure smooth convergence (see Appendix \ref{app : gradient at finite T} 
for further discussion and expressions for the analytic gradient of the cost function at finite temperature).
We matched the particle number on the impurity sites and on the lattice exactly 
by separately fitting the chemical potential using quadratic interpolation ~\cite{Zheng18}.
Direct inversion in the iterative subspace (DIIS)~\cite{Pulay80, Pulay82} was employed
to accelerate the overall DMET convergence,
using the difference of $u$ between two adjacent iterations as the error vector. We chose
the convergence threshold to be $10^{-4}$ in the correlation potential $u$ (per site), which
we observed to translate to an energy convergence per site of better than $10^{-4}$. We further analyze the numerical convergence and 
 error estimates for the DMET self-consistency in Appendix \ref{app: convergence}.

\subsubsection{Order parameters}
To characterize the doping dependence of the ground-state, we define average AFM and $d$-wave SC order parameters. As usual, the AFM order parameter is chosen as the staggered magnetization,  
\begin{equation}\label{eq: AFM order param}
m_{\AFM} = \frac{1}{4} \sum_{i\in \rm Cu} \eta^{\AFM}_i m^{d}_i,
\end{equation}
where $m^{d}_i$ is the local magnetic moment on a Cu-$d$ orbital, $\frac{1}{2} \qty(n^{d}_{i\alpha} - n^{d}_{i\beta})$, and the $\eta^{\AFM}$ is the local structure factor,
\begin{equation}\label{eq: eta AFM}
\eta^{\AFM}_i =\left\{
\begin{aligned}
&+1, ~ {\rm if} ~ n^{d}_{i\alpha} \geqslant n^{d}_{i\beta},  \\
&-1, ~ {\rm if} ~ n^{d}_{i\alpha} < n^{d}_{i\beta}.
\end{aligned}
\right.
\end{equation}   
The SC order parameter here is evaluated as the average of the Cu-Cu and O-O $d$-wave pairing components,
\begin{equation}\label{eq: SC order param}
\begin{split}
m_{\SC} &= \sum_{\expval{ii'}} \frac{1}{\sqrt{2}} \eta^{\SC}_{ii'} \qty(\expval{d_{i\alpha}d_{i'\beta}} + \expval{d_{i'\alpha}d_{i\beta}}) \\
&+  \sum_{\expval{\expval{jj'}}} \frac{1}{\sqrt{2}} \eta^{\SC}_{jj'}\qty(\expval{p_{j\alpha}p_{j'\beta}} + \expval{p_{j'\alpha}p_{j\beta}}),
\end{split}
\end{equation}
where $\expval{\cdots}$ limits the summation such that only the pairing between nearest Cu-$d$ orbitals is taken into account, and similarly $\expval{\expval{\cdots}}$ involves only the next-nearest coupling between O-$p$ orbitals. 
The $d$-wave superconducting structure factor $\eta^{\SC}$ is defined as,
\begin{equation}\label{eq: eta SC}
\eta^{\SC}_{ii'} =\left\{
\begin{aligned}
&+1, ~ {\rm if} ~ \vecR_{i} - \vecR_{i'} = \pm \vece_x ,  \\
&-1, ~ {\rm if} ~ \vecR_{i} - \vecR_{i'} = \pm \vece_y .
\end{aligned}
\right.
\end{equation}

\section{The three-band Phase diagram}\label{sec: results}

\subsection{Undoped state}\label{subsec: undoped results}

\begin{table}[!htb]
\caption{\label{tab:undoped}%
Charge, spin distribution (magnetic moments) and energy gap of the undoped three-band Hubbard model and reference data. 
Note that the experimental gaps reported are all optical gaps.}
\begin{ruledtabular}
\begin{tabular}{lccccc}
Model 			& $\rho_{\rmCu}$ & $\rho_{\rmO}$ & $m_{\rmCu}$ & $m_{\rmO}$ & $E_{\rm g}$ [eV]\\
\hline
Hybertsen       &  1.238 & 	1.881  &	0.363   & 0.000	 & 2.5 \\
Martin          &  1.219 &	1.891  &	0.375   & 0.001  & 4.4 \\
Hanke           &  1.220 &	1.890  &	0.373   & 0.000  & 3.9\\
Hanke full      &  1.358 &	1.821  &	0.279	& 0.002  & 2.2 \\
Others          &  1.23\footnotemark[1]      &  1.89\footnotemark[1]       &   0.29\footnotemark[2], 0.31\footnotemark[3]       &        & 2.25\footnotemark[2],  \\
Cuprate 		&		  &    	   &  $0.3 \pm 0.025$\footnotemark[4]        &        & 1.5-2.0\footnotemark[5], 1.5-1.7\footnotemark[6] \\ 
\end{tabular}
\end{ruledtabular}
\footnotetext[1]{DMRG result from Ref.~\cite{White15}, using a similar model to Hanke full (with a different $U_{p} = 4.5$, $V_{pd} = 1.5$ and $V_{pp} = 1.125$).}
\footnotetext[2]{VCA result from Ref.~\cite{Arrigoni09}, using basically the same model as Hanke full (with a different $U_{p} = 4.5$).}
\footnotetext[3]{VCA result from Ref.~\cite{Hanke10}, using the same model as Hanke full.}
\footnotetext[4]{Experimental result for \ce{La2CuO4}, from Ref.~\cite{Yamada87}.}
\footnotetext[5]{Experimental result for \ce{La2CuO4}, from Refs.~\cite{Tokura90, Cooper90-cuprate-gap, Uchida91}.}
\footnotetext[6]{Experimental result for \ce{YBa2Cu3O6}, from Refs.~\cite{Cooper90-cuprate-gap, Romberg90-YBCO-gap}.}
\end{table}

\subsubsection{Charge and magnetic moments} 
We present the order parameters for the undoped state from DMET and from reference calculations and experimental measurements in Table \ref{tab:undoped}.
As expected, the $d$ orbitals are roughly half-filled and the $p$ orbitals are roughly doubly occupied, with some charge transfer between the two
due to hybridization. 
Comparing the full and minimal parametrizations, in the full parametrization, the Cu site is more strongly occupied by electrons, due to the $t_{pp}$ term which smears out the oxygen charge and effectively transfers it to copper (while the effect of $U_{p}$ is very small,
see the discussion in Sec. \ref{subsec: doped results}).
Unlike on the O site, the spin density on the Cu site is polarized, with a large local magnetic moment, which compares well to the experimental value $0.3 \pm 0.025$~$\mu_{\rm B}$ ($0.6 \pm 0.05$ $\mu_{\rm B}$)~\cite{Yamada87}, as well as previously computed theoretical moments of 0.29~\cite{Arrigoni09} and 0.31~\cite{Arrigoni09} from VCA. In addition, the magnetic moment in the full parametrization is reduced relative
to the minimal parametrizations, because the increased electron density on copper dilutes the polarized spin, 
while the additional holes on oxygen reduce the strength of the super-exchange-based antiferromagnetic coupling. 
In fact, the local magnetic moments in the minimal models appear to be too large, while that of the full model is similar
to experimental results. However, it is also known from one-band calculations, that the magnetic moments are overestimated
in $2\times 2$ DMET clusters relative to the thermodynamic limit (e.g., by about 25\% at $U=6$, see Fig. S1 in the Supplemental
Material~\cite{note-SI-3band}. Assuming similar finite size errors, then the minimal parametrization may provide reasonable magnetic moments at half-filling in the thermodynamic limit (although this is not necessarily
the case under doping, see below).

\subsubsection{Band gap} 
As a simple estimate of the single-particle gap, we also computed the energy gap of the converged DMET non-interacting lattice
Hamiltonian (DMET NI gap), i.e. $E_{\rm g} = \varepsilon_{\rm CBM} - \varepsilon_{\rm VBM}$, where C(V)BM denotes conduction (valence) band minimum (maximum).
Note that although the charge and spin densities in the different
parametrizations are generally similar, the DMET NI gap varies more significantly, from 2.2 to 4.4 eV. The Hybertsen and Hanke parameter
sets were derived from calculations on \ce{La2CuO4} (LCO), where the optical energy gap is variously reported as lying in the
range 1.5 to 2.0 eV ~\cite{Tokura90, Cooper90-cuprate-gap, Uchida91} (note that
the optical gap is generally smaller than the fundamental gap). 
The estimated DMET NI gap of 2.5 and 2.2 eV for the Hybertsen and Hanke full parameter set
respectively are thus in reasonable agreement with the experimental gap.
However, the minimal Hanke parametrization seriously overestimates the gap. The Martin parameter set, obtained from calculations
on finite-sized Cu-O clusters, are all systematically larger than in the other sets, and thus give the largest DMET NI gap. However, since
the ratio of parameters in the Martin model remains similar to other parametrizations (and thus give rise to similar charge and spin distributions)
this suggests that all energy parameters in the Martin model should simply be simultaneously rescaled downwards.

\begin{figure*}[!htb]
\includegraphics[width=160mm, clip]{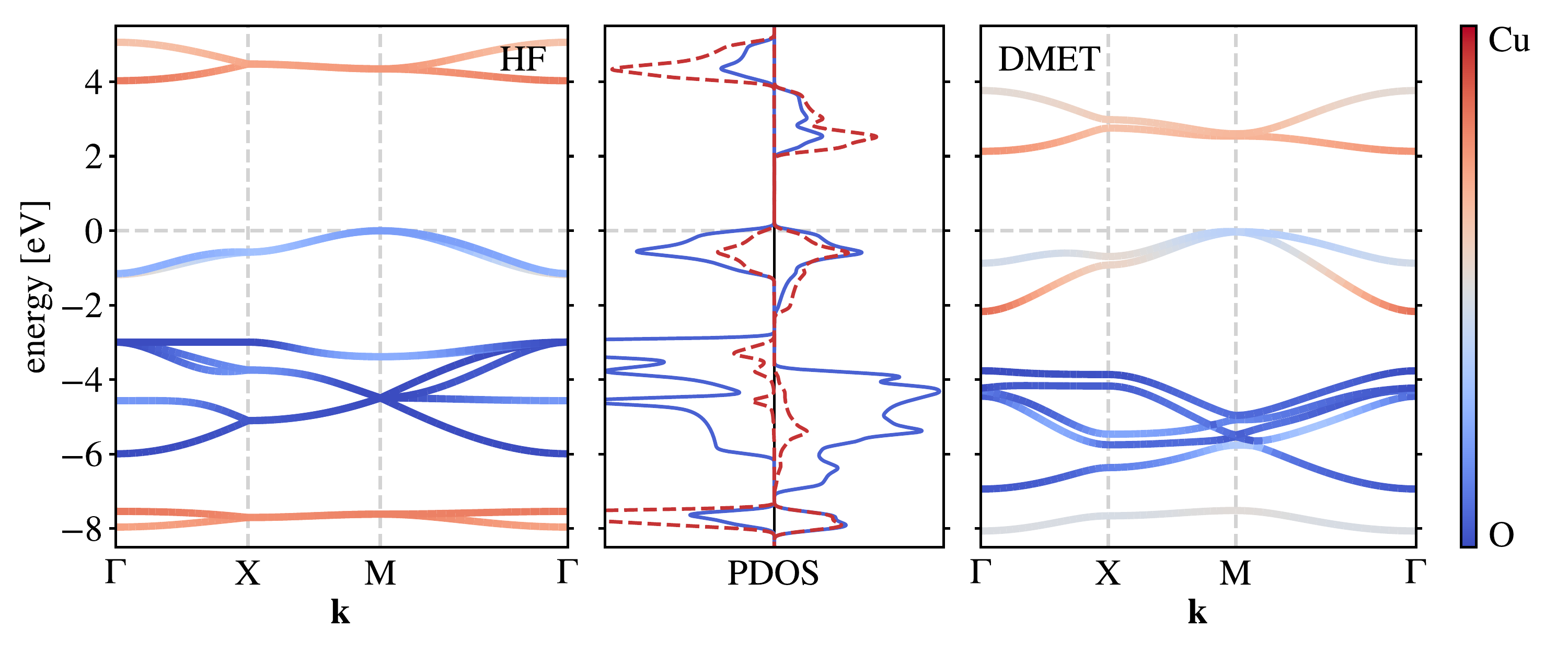}
\caption{\label{fig:HF DMET bands} Orbital-projected electronic band structure and density of states (PDOS) of the undoped three-band Hubbard model with Hanke full parameters from HF (left) and DMET (right). The special $\veck$ points [$\Gamma$: $(0, 0)$, X: $(\pi, 0)$, M: $(\pi, \pi)$] are in the first Brillouin zone of the $2\times 2$ supercell lattice. The valence band maximum (VBM) is chosen as the energy zero. }
\end{figure*}

\subsubsection{Orbital resolved band structure} 
Unlike in the one-band Hubbard model, where the insulating gap arises between Hubbard bands,
the gap in correlated insulators in the three-band model can arise from both Hubbard and charge-transfer mechanisms.
In Fig. \ref{fig:HF DMET bands}, we plot the projected electronic band structure and density of states from the DMET non-interacting lattice Hamiltonian, as converged for the fully parametrized Hanke model. The CBM is mainly of Cu $d$ character (upper Hubbard band),
while the VBM shows mixed character, dominated somewhat by O-$p$.
The mixed orbital character of the valence bands around the Fermi level
is consistent with the Zhang-Rice singlet (ZRS) hypothesis~\cite{Zhang88ZRS}, in which hybridization between oxygen and
copper orbitals induces superexchange that leads to singlets of O and Cu holes.
Further support for the ZRS picture comes from the $\veck$-dependent orbital weights; that of Cu-$d$ is greater at the $\Gamma$ point, while that of  O-$p$ is larger at the M point, consistent with earlier model analysis of the ZRS state~\cite{Jefferson92} and results from VCA~\cite{Arrigoni09}.
In total, these observations indicate that the undoped three-band model ground state is a charge transfer insulator, with mainly a $p$-$d$ type energy gap (see Ref. ~\cite{Damascelli03} for experimental evidence of the charge-transfer nature of the band gap). The strong $\veck$-dependent
hybridization clearly poses challenges for numerical downfolding techniques to a one-band picture.

Comparing the DMET NI band structure to the Hartree-Fock mean-field description (also shown in Fig.~\ref{fig:HF DMET bands}), we find that the HF
gap ($\approx 4$ eV) is significantly overestimated, and the $d$-$p$ hybridization is significantly weaker, resulting in a
VBM with dominant oxygen $p$ character and very narrow dispersion. Thus the reduced gap and $d$-$p$ hybridization, both seen in experiment,
are fluctuation driven phenomena, whose average effect is being captured by the DMET correlation potential $u$.

\subsection{Doped states}\label{subsec: doped results}

\subsubsection{Hole doped phases with standard parametrizations}
More interesting ground states, including those with superconducting order, appear under doping.
An important difference with the one-band case is the asymmetry of the three-band model under doping. We
first focus on the orders that appear under hole-doping.
Although our calculations are all at zero temperature, 
we can loosely identify the magnitude of the order parameters with transition temperatures in the phase diagram,
thus allowing us to compare them to the experimental phase diagram.
\begin{figure}[!htb]
\center
\includegraphics[width=85mm, clip]{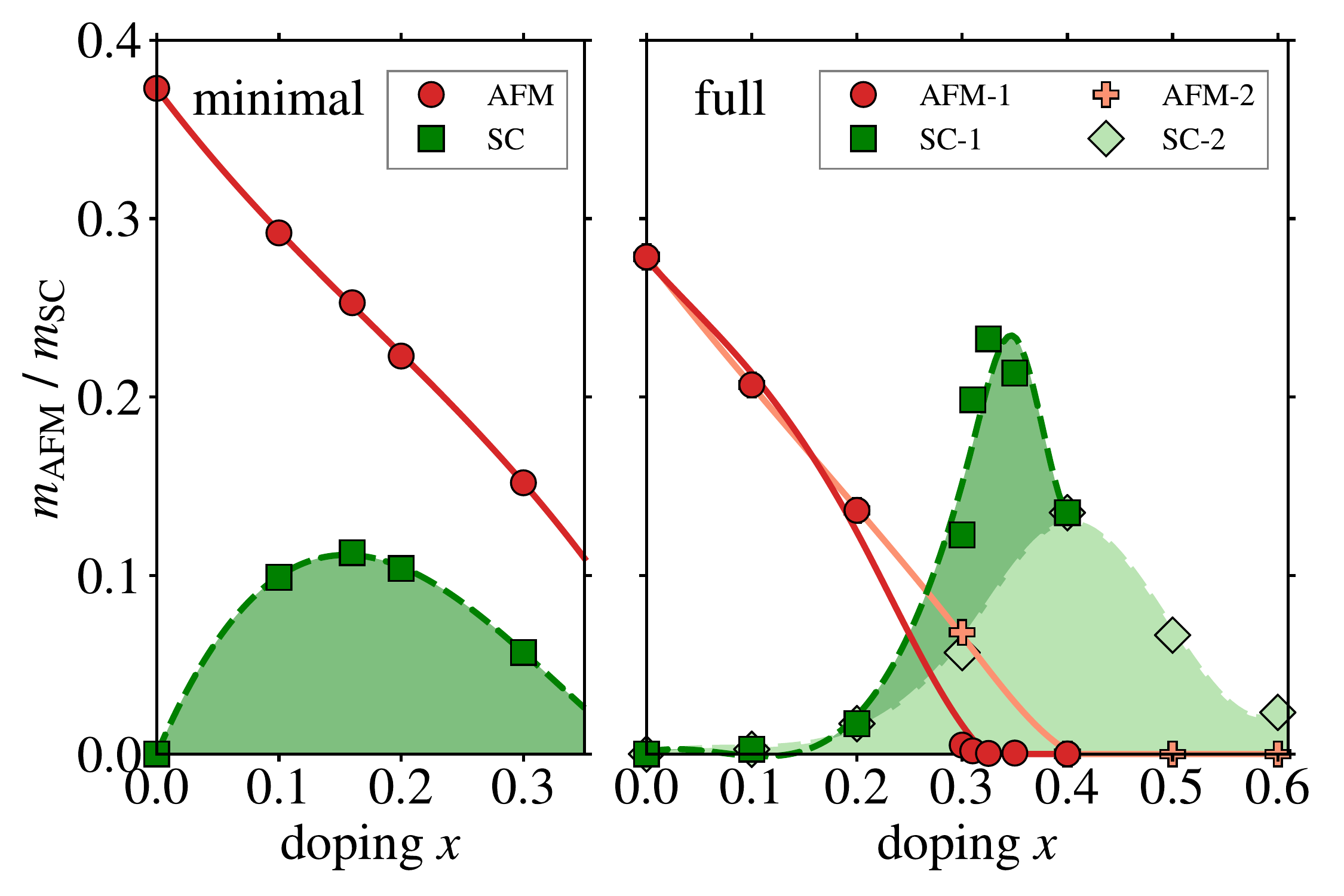}
\caption{\label{fig:order-param} Antiferromagnetic and $d$-wave superconducting
	order parameters of the hole-doped three-band Hubbard model. The model
	settings are from the Hanke minimal (left) and Hanke full (right) parameter
	sets. Note that in the ``Hanke full'' case, we find
	two possible solutions between $x = 0.2$ and $x = 0.4$, marked as solution
	``1'' (from a weakly polarized AFM guess) and ``2'' (from a strongly
	polarized AFM guess) in the figure. The curves are cubic-spline interpolated.}
\end{figure}
In Fig. \ref{fig:order-param}, we plot the AFM and $d$-wave SC order parameters of the Hanke model as a function of hole doping (Hybertsen and Martin minimal model results are very similar to those of the Hanke minimal model, as shown in Fig. S2 of the Supplemental Material~\cite{note-SI-3band}). In the fully parametrized model, we find two different solutions of the DMET self-consistency, 
labelled solution 1 (obtained from a weakly spin polarized AFM guess) and solution 2 (obtained
from a strongly polarized AFM guess). 



For all parameter sets, we observe that the AFM order parameter decreases as doping increases, consistent with the general
behavior of the cuprate phase diagram ~\cite{Damascelli03}.
However, for the minimal models, the AFM order persists even up to large dopings (e.g. $\approx 0.15$ at $x = 0.3$).
In interpreting this discrepancy, one complication is that computation is measuring atomic scale local order, while
experimental measurements are likely averages over various inhomogeneities (e.g. different orientations of stripes in different layers) which
would typically lead to reduced moments.
Leaving this aside, however, the overestimation of the computed moment could originate either from the remaining finite size error in the DMET calculation, or
from the unphysical nature of the parametrization (e.g. the lack of doping dependence of the parameters). From our earlier work
on the one-band Hubbard model~\cite{Zheng16}, we
know that DMET calculations using a $2\times 2$ impurity (e.g. in the range $U / t = 6 - 8$) indeed overmagnetize not only
at half-filling but also in the doped regime (see Fig. S1 of the Supplemental Material~\cite{note-SI-3band}). 
However, the one-band AFM order nonetheless vanishes at dopings larger than $0.25$, more rapidly
than what we observe in the minimal parametrized three-band model. In addition, the full parametrization of the three-band model also
predicts a more realistic trend for the AFM order at large doping.
Taken together, this suggests that the observed persistent AFM order is likely due to the oversimplified
minimal model parameters.
 Although the AFM order in the full model does decrease to zero in the observed
doping range, it vanishes between $x = 0.2$ and $0.3$ (more similar to the one-band model). This is beyond the experimental boundary
for the pure AFM phase ($x < 0.1$), but close to the boundary of the pseudogap region~\cite{Timusk99, Lee06RMPhightc}. Like in the one-band
model, we would expect longer wavelength orders (such as striped phases~\cite{Zheng16, Zheng17sci}) to appear in this region with larger computational clusters. 

From Figs. \ref{fig:order-param} and S2, we see that $d$-wave superconducting order (coexisting with antiferromagnetism) appears
in the phase diagram of all  parameter sets. (Discussion of additional pairing orders, as well as comparisons to the one-band model
can be found further below). 
In the minimal models, the $d$-wave pairing  reaches a maximum at around $x= 0.15 - 0.20$.
As a result of the overestimation of AFM order discussed above, the minimal models
show coexistence of AFM + SC order for all the studied dopings.  However, in the full parametrization, the two coexist in the range $0.1$ to $0.4$ (for solution 2), and  $0.1$ to $0.3$ (for solution 1), with $d$-wave order
reaching a maximum near $x \approx 0.30 - 0.35$, somewhat
larger than seen in experiments ($\approx 0.15$ - $0.2$)~\cite{Yamada87}. Solutions 1 and 2 coincide for $x < 0.2$ and $x>0.4$ but are distinct in between, 
reflecting the known competition between orders at intermediate doping ~\cite{Lee06RMPhightc}; solution 1 is slightly lower in energy
and displays significantly stronger superconducting order. Note that it is also possible to converge a paramagnetic SC solution (by constraining
the correlation potential in Eq. \ref{eq: u potential} so that $v^{\alpha} = v^{\beta}$ and $\Delta = \Delta^{\dg}$). In this case, the SC order is already evident at $x = 0.1$, since the AFM order is artificially suppressed. However, the energy of this paramagnetic state is much higher than the AFM + SC states we have
discussed, and is unstable if one releases the constraints on the potential. We thus believe the coexistence of AFM and SC order
to be a true feature of the three-band model ground state, as has also been
observed in VCA studies ~\cite{Arrigoni09, Hanke10}.

\begin{figure}[!htb]
\includegraphics[width=85mm, clip]{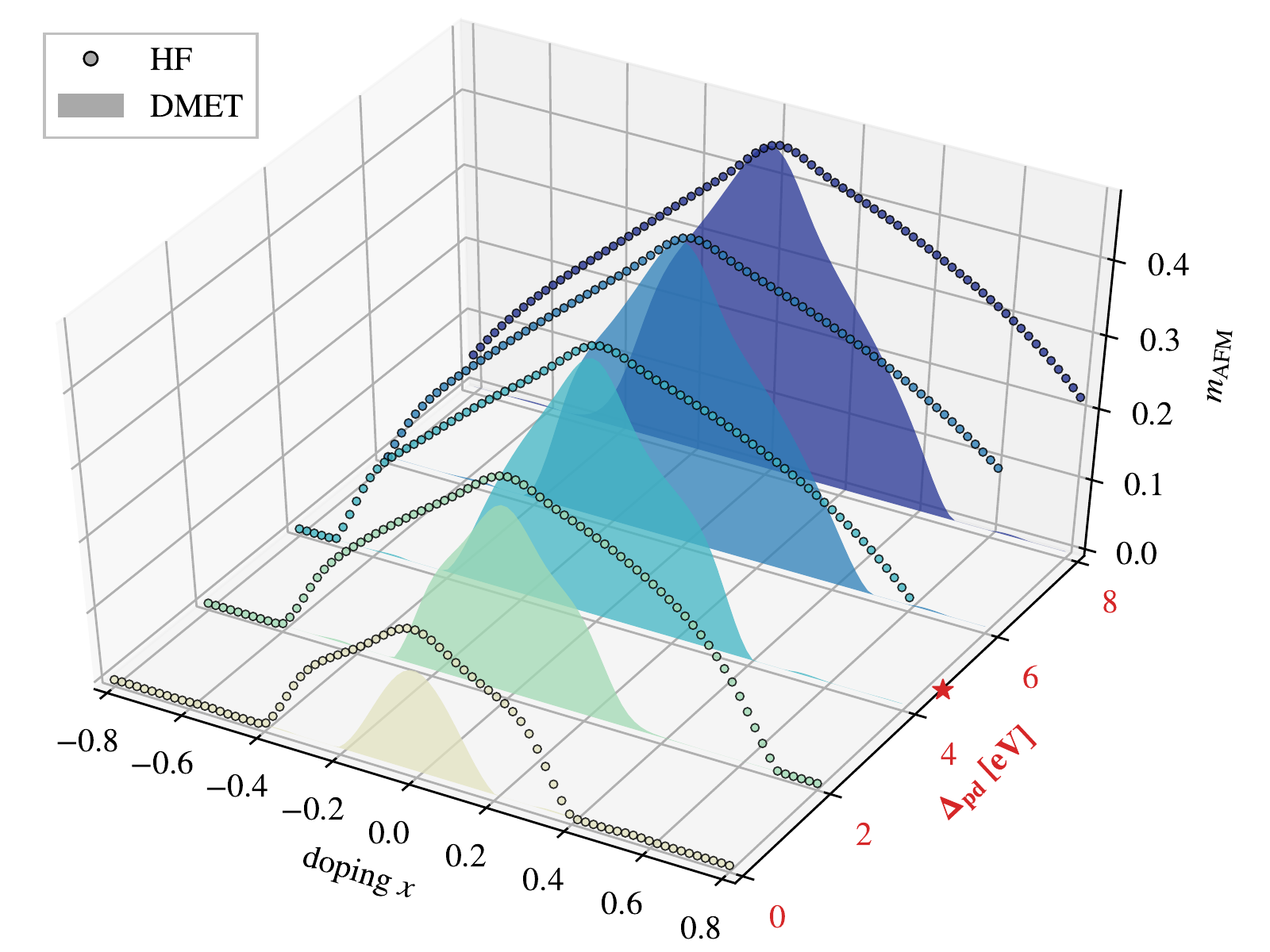}
\caption{\label{fig: m-vs-Dpd}Effects of $\Delta_{pd}$ on the magnetic phase diagram of the three-band Hubbard model. $\Delta_{pd}$ ranges from 0.0 to 8.0 eV and the star marker labels the value used in the Hanke model (4.5 eV). Both Hartree-Fock (dotted line) and DMET (shaded area) results are shown.}
\end{figure}

\begin{figure}[!htb]
\includegraphics[width=85mm, clip]{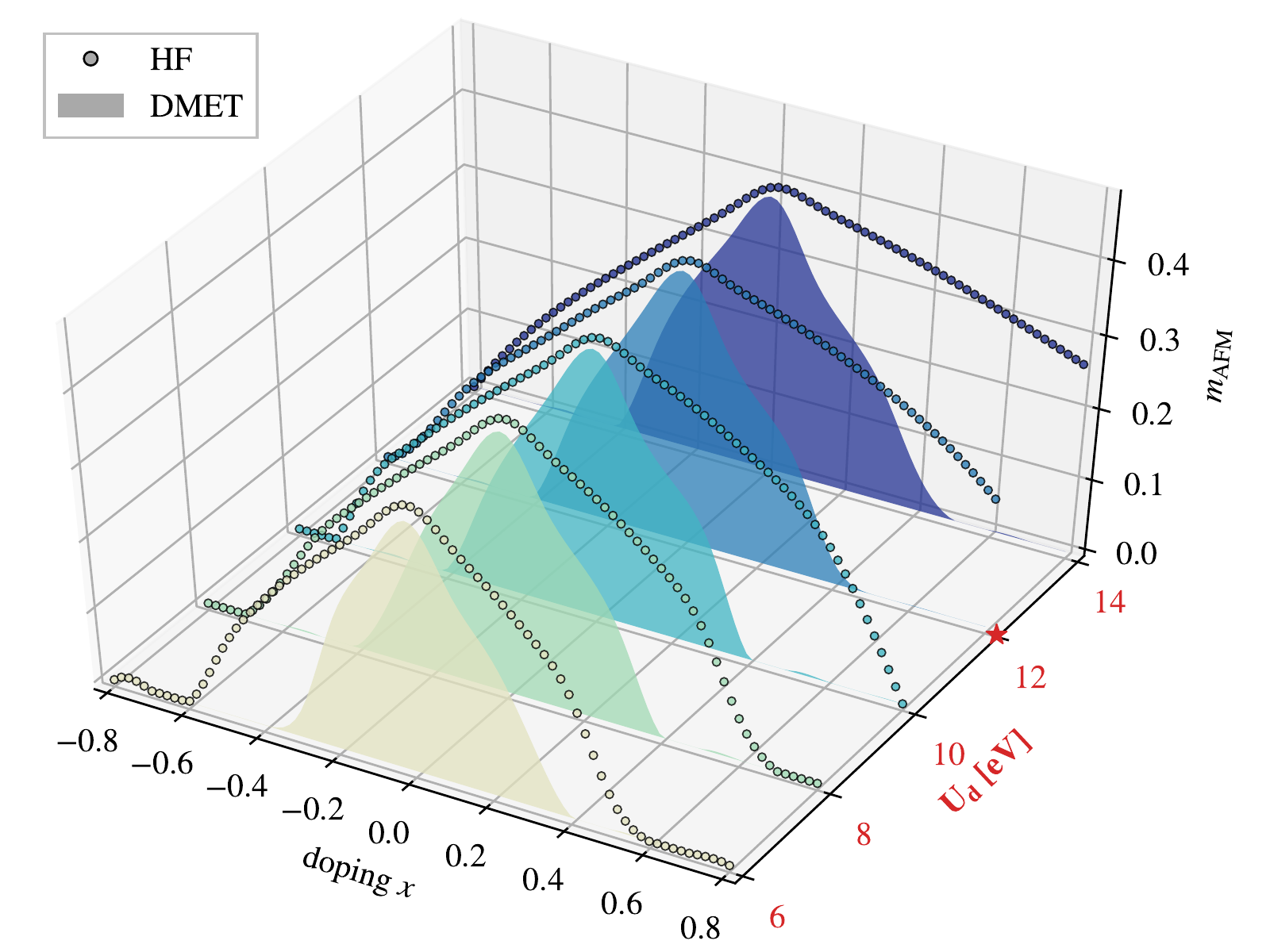} 
\caption{\label{fig: m-vs-Ud}Effects of $U_{d}$ on the magnetic phase diagram of the three-band Hubbard model. $U_{d}$ ranges from 6.0 to 14.0 eV and the star marker labels the value used in the Hanke model (12.0 eV). Both Hartree-Fock (dotted line) and DMET (shaded area) results are shown.}
\end{figure}

\begin{figure}[!htb]
\includegraphics[width=85mm, clip]{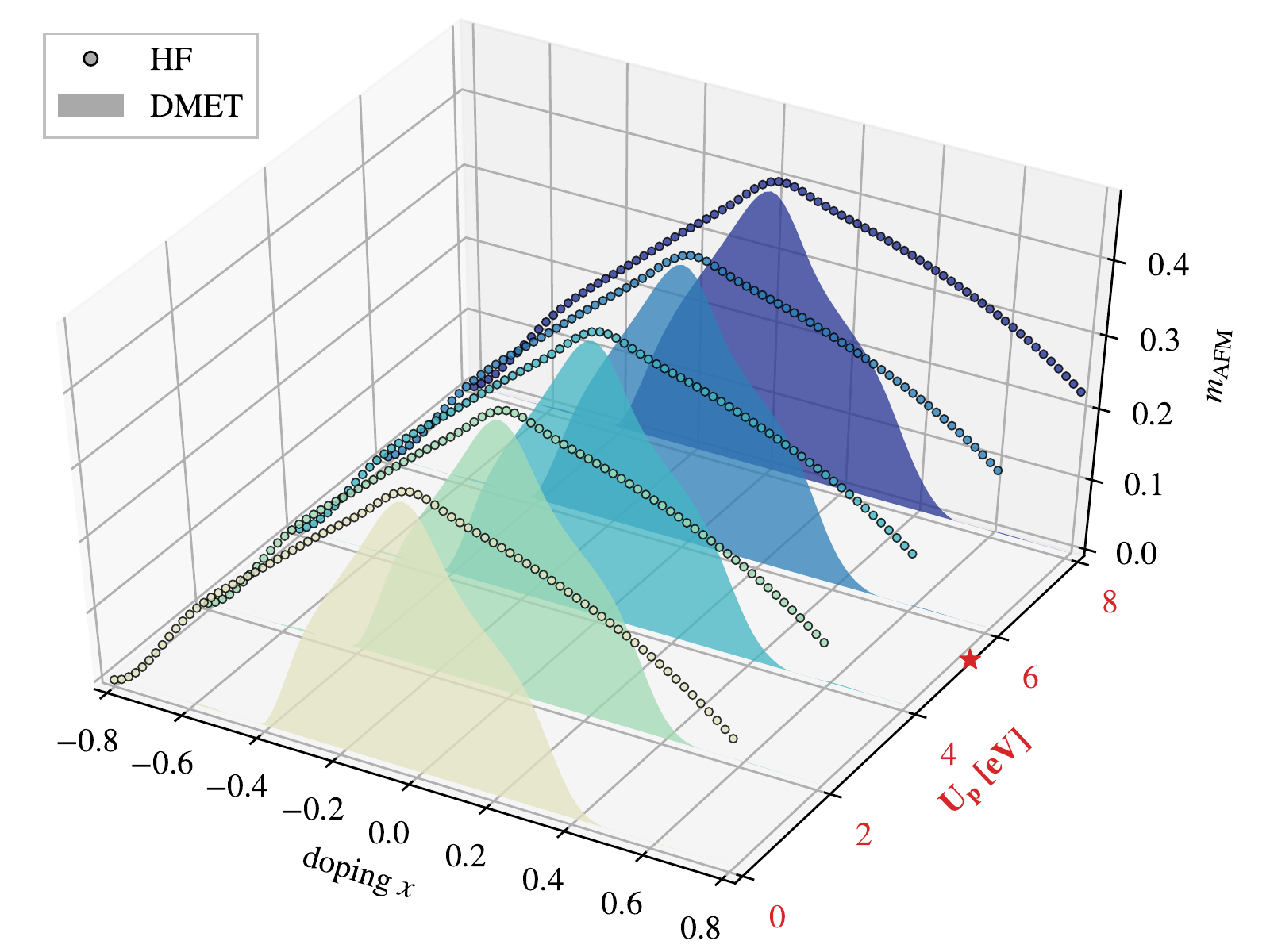} 
\caption{\label{fig: m-vs-Up}Effects of $U_{p}$ on the magnetic phase diagram of the three-band Hubbard model. $U_{p}$ ranges from 0.0 to 8.0 eV and the star marker labels the value used in the Hanke full model (5.25 eV). Both Hartree-Fock (dotted line) and DMET (shaded area) results are shown.}
\end{figure}

\begin{figure}[!htb]
\includegraphics[width=85mm, clip]{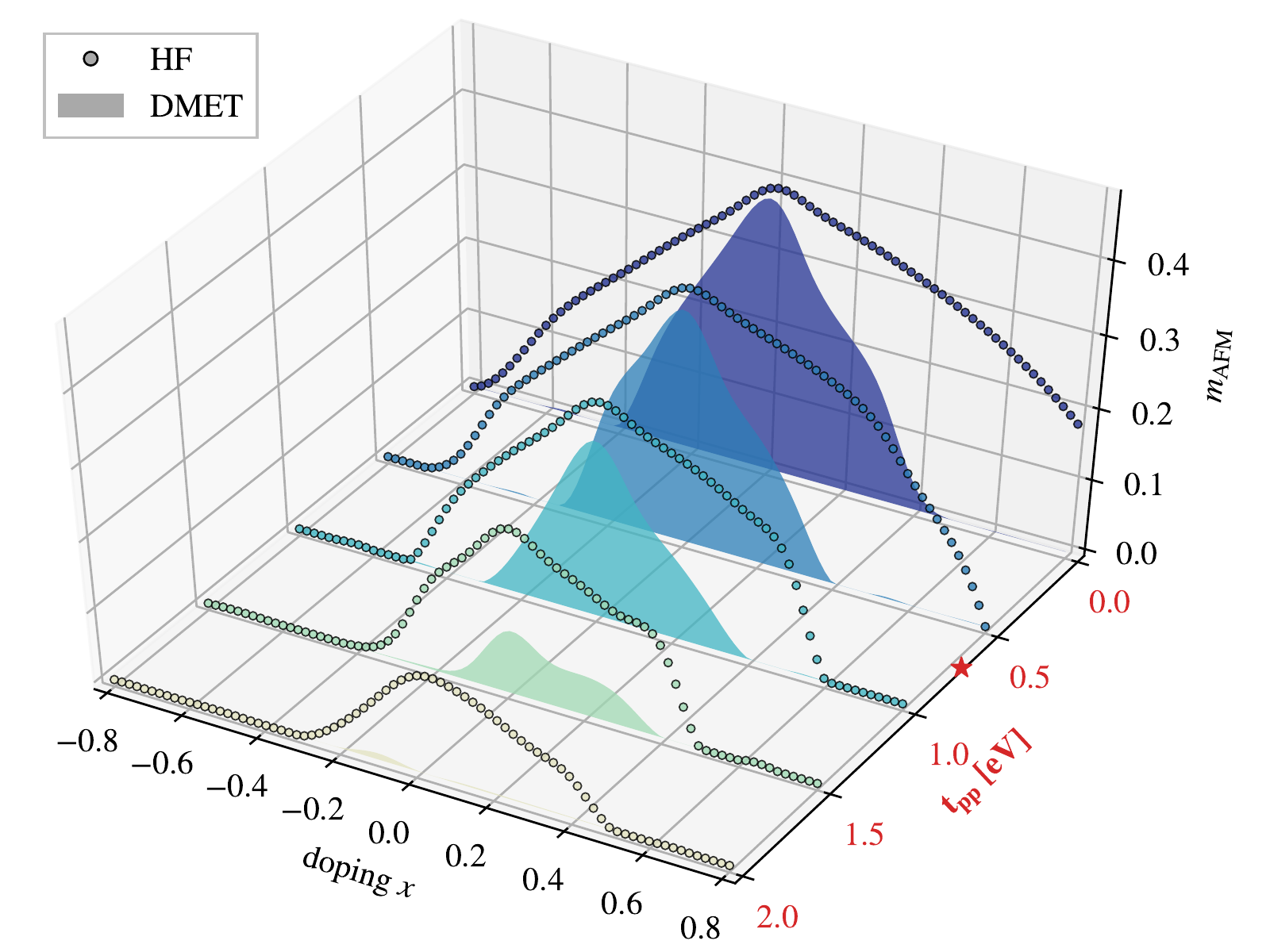} 
\caption{\label{fig: m-vs-tpp}Effects of $t_{pp}$ on the magnetic phase diagram of the three-band Hubbard model. $t_{pp}$ ranges from 0.0 to 2.0 eV and the star marker labels the value used in the Hanke full model (0.75 eV). Both Hartree-Fock (dotted line) and DMET (shaded area) results are shown.}
\end{figure}

\subsubsection{Range of reasonable parameters} 
In view of the significant differences between the minimal and full parametrizations, we now
examine more deeply how individual parameters influence the phase diagram. To do so, we change the individual parameters appearing in the
Hanke minimal model, and restrict ourselves to the magnetic order for simplicity. We compare the magnetic phase diagram computed
using both Hartree-Fock and DMET. While the mean-field Hartree-Fock method overestimates the magnetic moments, and the resulting AFM domes
always lie above the DMET ones in the plots, it should be noted that parametrizations are often derived from mean-field calculations.
Thus the difference in sensitivity between DMET and HF to the model parameters gives some insight
into the sizes of errors arising from mean-field parametrization schemes.

We first study the influence of $\Delta_{pd}$, whose value is uncertain in the literature \cite{Kent08, Chiciak18};
the magnetic phase diagram computed using both Hartree-Fock and DMET is shown in Fig. \ref{fig: m-vs-Dpd}.
On the hole-doped side, when $\Delta_{pd} \geqslant 4$ eV, the HF magnetic moment does not vanish even at a large doping of $x \approx 0.8$, while in contrast, DMET always predicts a finite AFM  region with a sharp peak at $x = 0$.
The DMET magnetic moment $m_{\AFM}$  increases monotonically from 0.14 to 0.44 as we increase $\Delta_{pd}$, which can be understood from second-order perturbation theory: The effective $d$-$d$ hopping $t_{dd} \propto \frac{t^2_{pd}}{\Delta_{pd}}$, thus a larger $\Delta_{pd}$ gives a smaller $t_{dd}$ and thus enhances the magnetic moment. Along with the
larger moments, the critical doping point where $m_{\AFM} \rightarrow 0$  shifts to larger doping as $\Delta_{pd}$ increases.
Given that, even accounting for finite cluster errors (see above), the minimal model appears
to overestimate the magnetic moment under doping and the critical doping concentration,
these results suggest one should renormalize $\Delta_{pd}$
to smaller values, around 2-3 eV. Finally, we see that the asymmetry with respect to electron and hole doping becomes more pronounced
when $\Delta_{pd}$ increases, and the magnetic moment is less sensitive to hole doping rather than  electron doping. Thus, the
appropriate value of $\Delta_{pd}$ should neither be too small (as the AFM order as well as doping asymmetry will both be too weak, see also Ref. \cite{Kent08} for a discussion of the unphysical behavior with small $\Delta_{pd}$) nor too large ($m_{\AFM}$ order will be too strong to be suppressed by doping, especially on the hole doped side). Ref. ~\cite{Weber12} suggests a range (1.2 - 2.6 eV) of $\Delta_{pd}$ for cuprates, which overlaps
 the range of our estimates.

We next check the effect of on-site Coulomb repulsion terms. The moment versus $U_{d}$ is shown in Fig. \ref{fig: m-vs-Ud}. Unlike $\Delta_{pd}$, the influence of $U_{d}$ on the shape of the curves is very small, e.g. the undoped DMET $m_{\AFM}$ only increases from 0.35 to 0.38 when $U_{d}$ varies from 6 to 14 eV. The influence on the curve shape is more significant for HF than it is for DMET. In the one-band Hubbard model, however, the situation is very different, where  $m_{\AFM}$ increases substantially as $U$ is increased~\cite{Zheng16}. This observation supports viewing the three-band model as primarily a \textit{charge transfer insulator} (and thus less sensitive to the change in the on-site Coulomb $U_{d}$), rather than a Mott insulator, whose magnetic moment is directly mediated by $U$. The situation for the on-site Coulomb repulsion $U_{p}$ (see Fig. \ref{fig: m-vs-Up}) is very similar to that for $U_{d}$:
the undoped DMET $m_{\AFM}$ only increases from 0.37 to 0.38 as $U_{d}$ varies from 0 to 8 eV, and the HF curves
show a similarly weak sensitivity.

We finally study the effect of nearest neighbor oxygen hopping $t_{pp}$ (see Fig. \ref{fig: m-vs-tpp}). From the figure, we see that the AFM order is effectively frustrated by large $t_{pp}$, similar to the effect of $t'$ in the one-band Hubbard model. It has been shown in Ref. ~\cite{Kent08} that $t_{pp}$
can vary substantially for different cuprates (unlike $t_{pd}$, which is almost unchanged between materials). Our results here suggest 
that a reasonable range for this parameter is around 0.5 - 1.0 eV; too large a $t_{pp}$ will suppress the AFM order.

Overall, we find that the magnetic phase diagram is sensitive to $\Delta_{pd}$ and $t_{pp}$, but not to $U_{d}$ and $U_{p}$. The improved results of the full model are thus likely due to the introduction of $t_{pp}$, rather than $U_{p}$. In particular, if we wish to have a reasonable description of the three-band Hubbard model within a minimal set of  parameters, $\Delta_{pd}$ should be renormalized to a smaller value to take the effect of $t_{pp}$ into account.  The Hanke parametrization of the full model yields more physical results, and thus we will only use
this full model in the remainder of the discussion. However, we note that it is still not optimal
with respect to choosing values of $\Delta_{pd}$ and $t_{pp}$ that match experiment. This may in part be due to the mean-field 
derivation of some of the parameters. 

\subsubsection{Electron doped phases in the full model} 
\begin{figure}[!htb]
\includegraphics[width=85mm, clip]{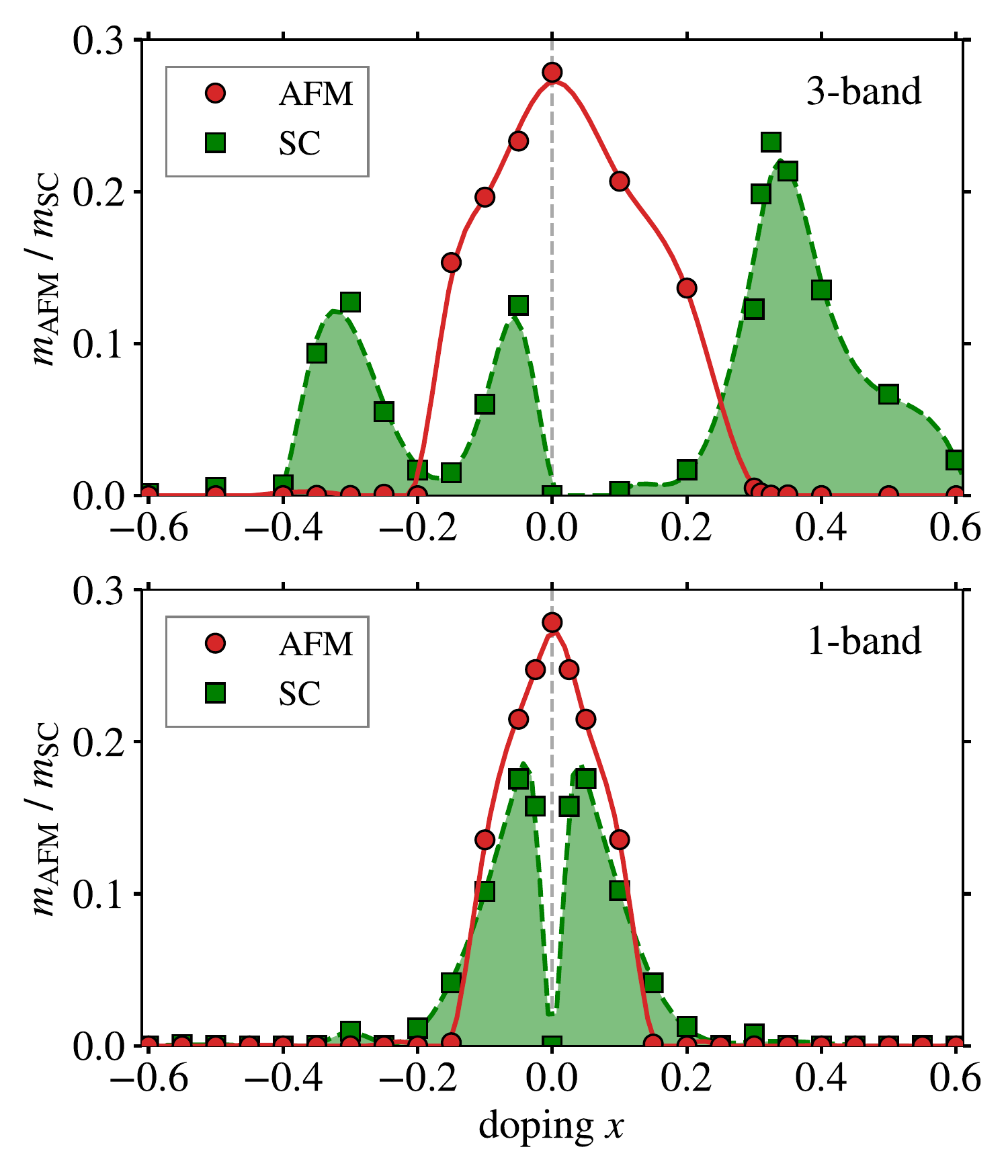} 
\caption{\label{fig: elec-1band}Comparison of electron doped ($x < 0$) and hole doped ($x > 0$) orders. Upper panel: AFM and SC order of the three-band Hubbard model (Hanke full parameter set). Lower panel: AFM and SC order of the one-band Hubbard model ($2\times2$ DMET cluster), with $U$  fitted such that at $x = 0$, $m_{\mathrm{AFM}}$ is the same as that of the three-band model.}
\end{figure}

We now turn to
the electron doped orders, which as mentioned above,
are different from the hole-doped orders, unlike in the one-band model~\cite{Lee06RMPhightc, White15}. We show the AFM and SC order versus both
hole and electron doping in
Fig. \ref{fig: elec-1band} (the hole doped side corresponds
to solution 1 in Fig. \ref{fig:order-param}). As we dope with more
electrons, the AFM order diminishes. The
critical doping $x_{\mathrm{c}}$ that makes $m_{\AFM}$ vanish (0.15 - 0.20) is smaller
than that on the hole doped side. This is quite different from what is seen in experiment: 
the commonly accepted cuprate phase diagram
typically shows a sudden drop of AFM order on the hole doped side
\cite{Damascelli03}, with a larger region of coexistence on the electron doped side.
This likely reflects the fact that a single parameter set does not
describe the electron-doped and hole-doped materials equally well.

For the SC phase, the overall $d$-wave pairing magnitude is smaller in the electron doped region, similar to the lower $T_{\mathrm{c}}$s seen in experiment.
Also, the SC phase on the electron doped side has an interesting ``M'' shaped two-peak structure: The $d$-wave SC order increases first with respect to the doping, but decays to a small value around the AFM critical $x_{\mathrm{c}}$, before growing to another peak after the AFM order vanishes. The first peak around $x = 0.05$ is very similar in shape to the peak in DMET calculations of the one-band Hubbard model, where the SC order emerges immediately
after doping (see the lower panel of Fig. \ref{fig: elec-1band}). The second peak, occurring after the disappearance of the AFM order,
is similar to the hole doped SC peak. The presence of two qualitatively different SC phases may be a hint
of the types of competing orders that can arise on the electron-doped side, which to date have not been much investigated
in numerical studies.


\begin{figure*}[!htb]
\subfigure{\label{fig:pattern a}\includegraphics[width=40mm, clip]{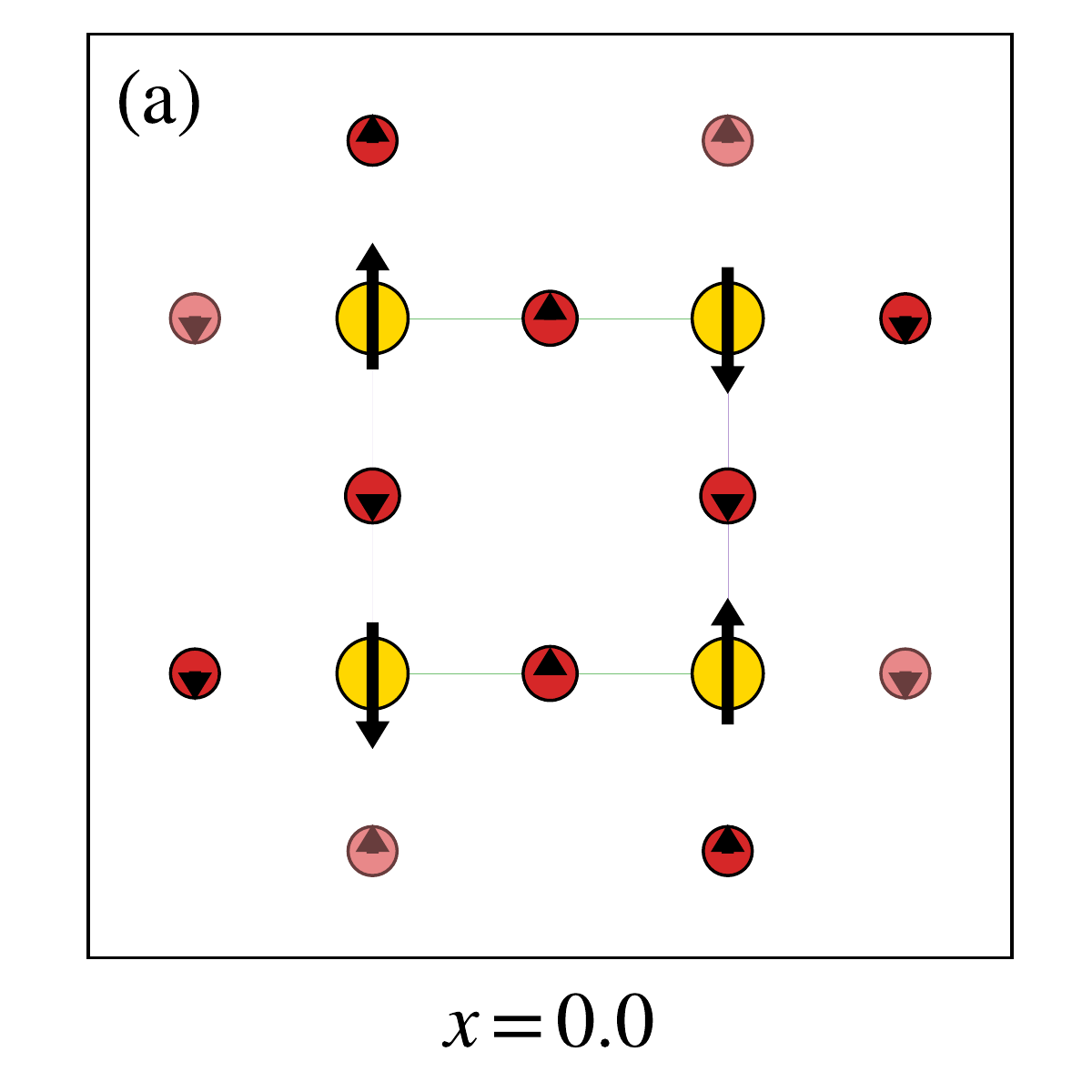}}
\subfigure{\label{fig:pattern b}\includegraphics[width=40mm,  clip]{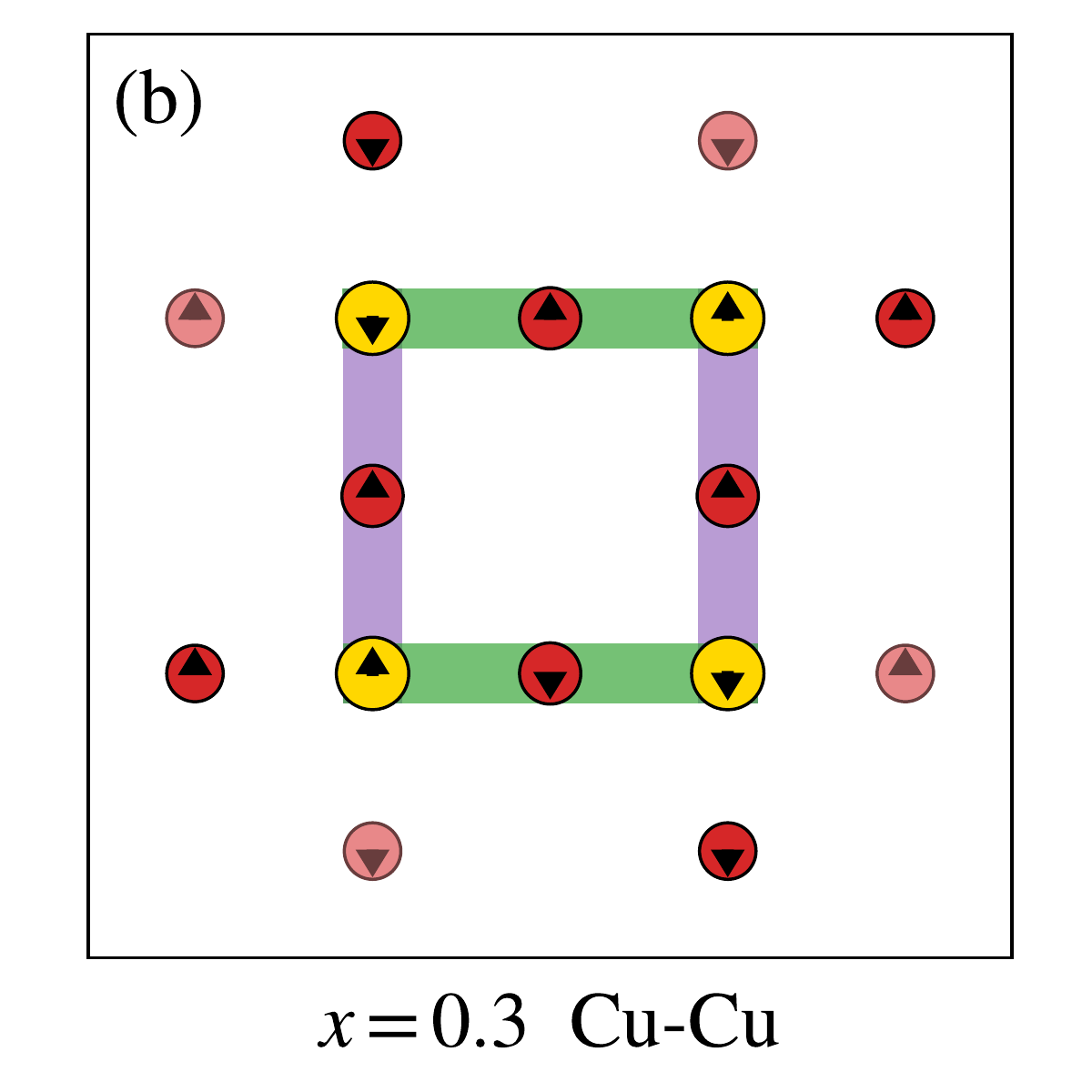}}
\subfigure{\label{fig:pattern c}\includegraphics[width=40mm,  clip]{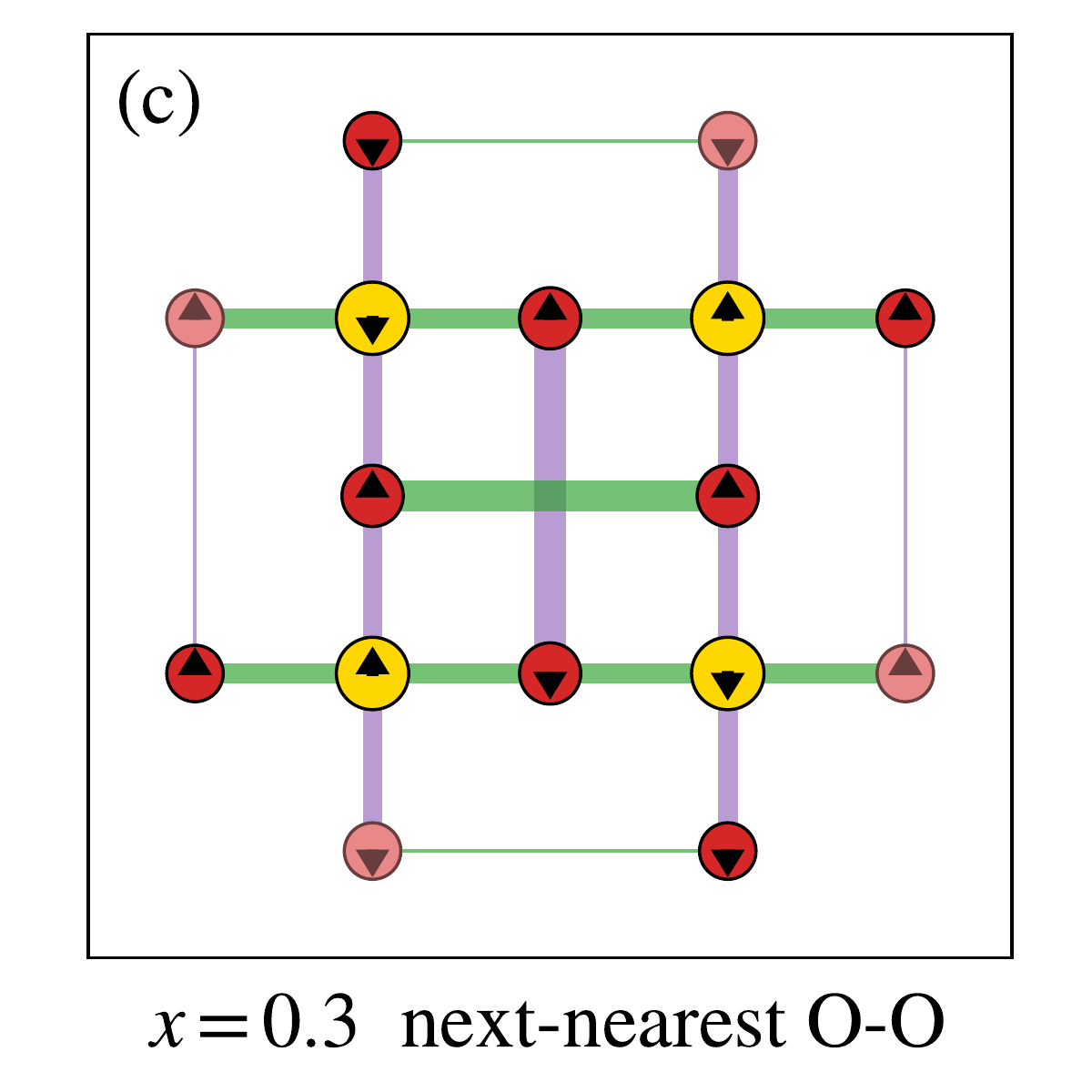}}
\subfigure{\label{fig:pattern d}\includegraphics[width=40mm,  clip]{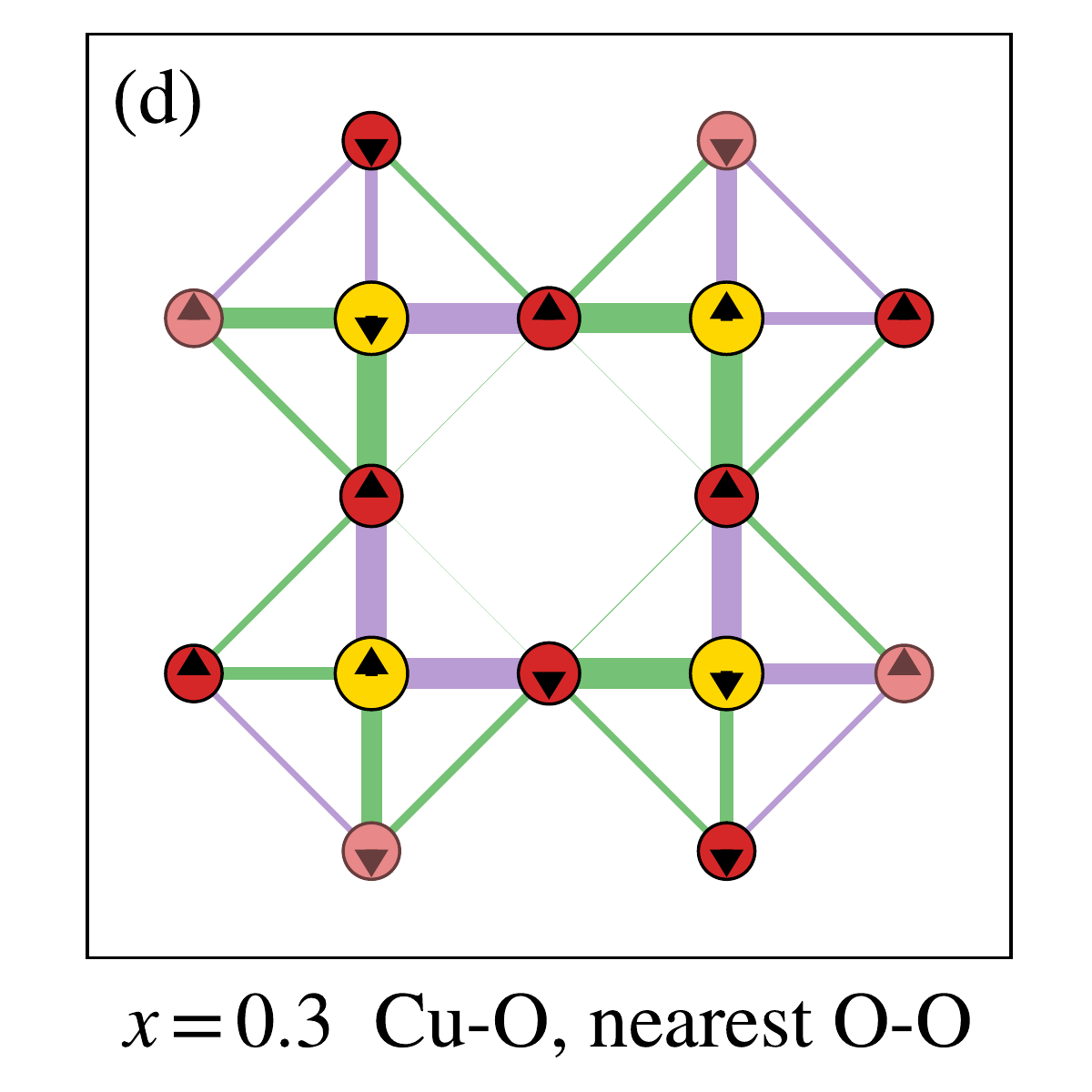}}

\subfigure{\label{fig:pattern e}\includegraphics[width=40mm,  clip]{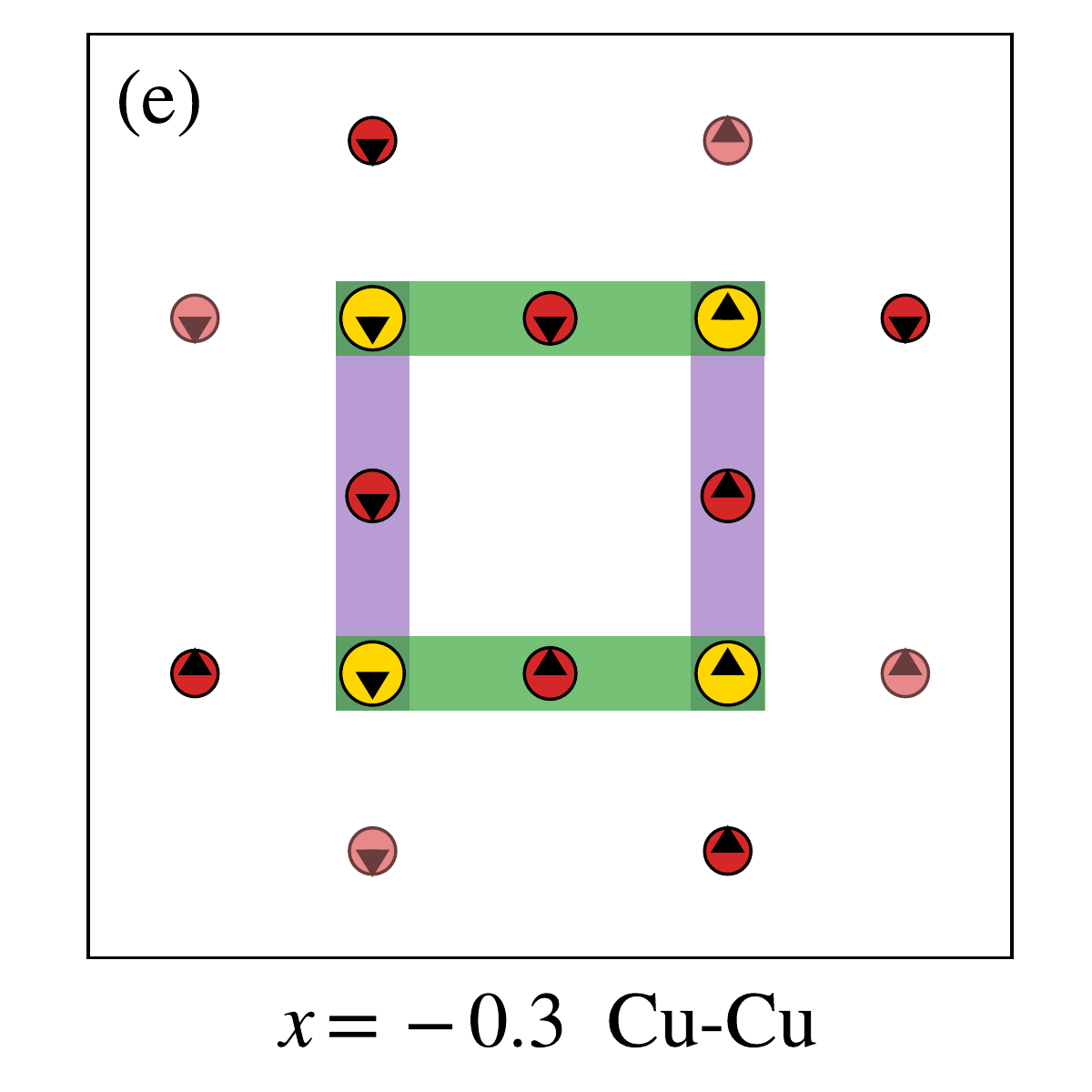}}
\subfigure{\label{fig:pattern f}\includegraphics[width=40mm,  clip]{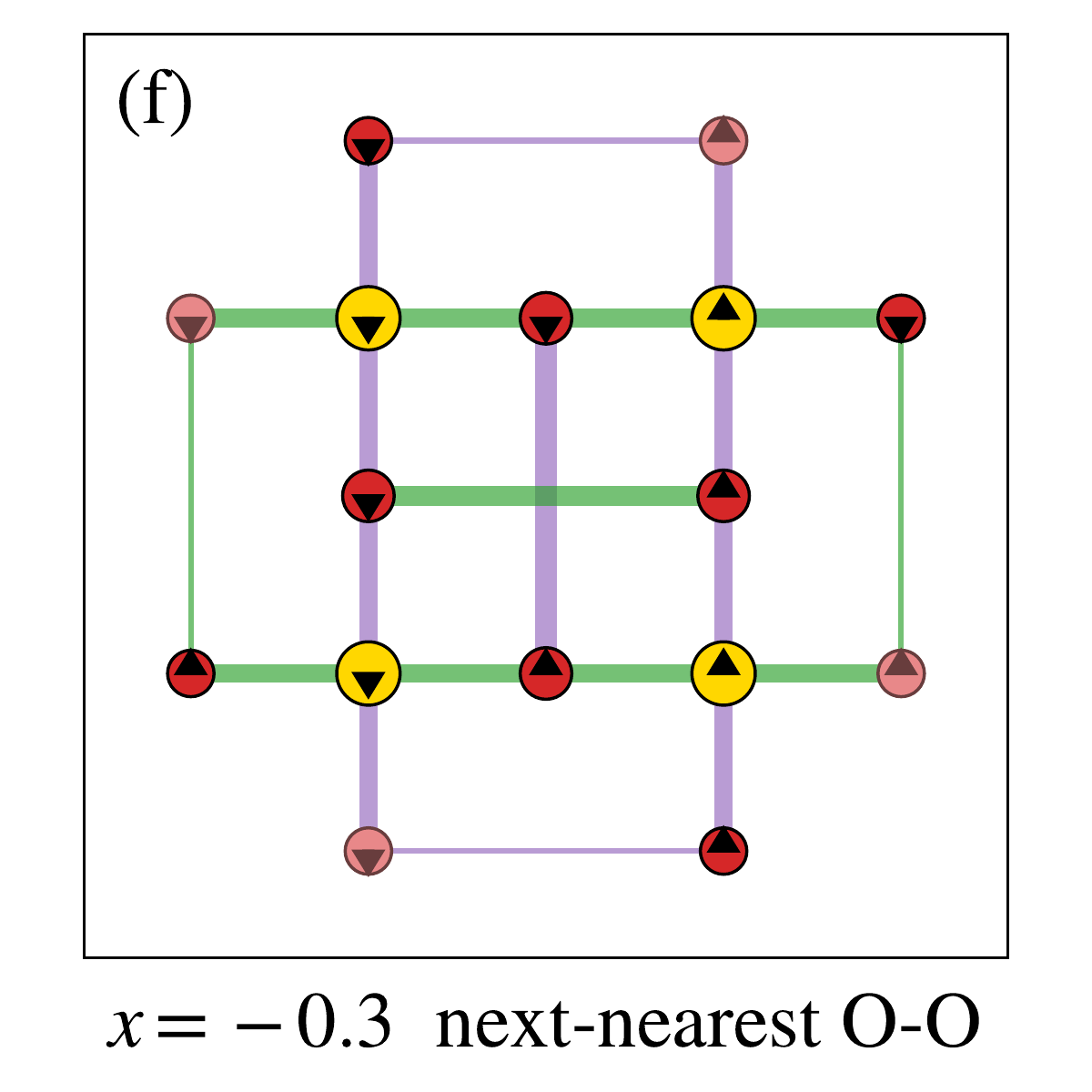}}
\subfigure{\label{fig:pattern g}\includegraphics[width=40mm,  clip]{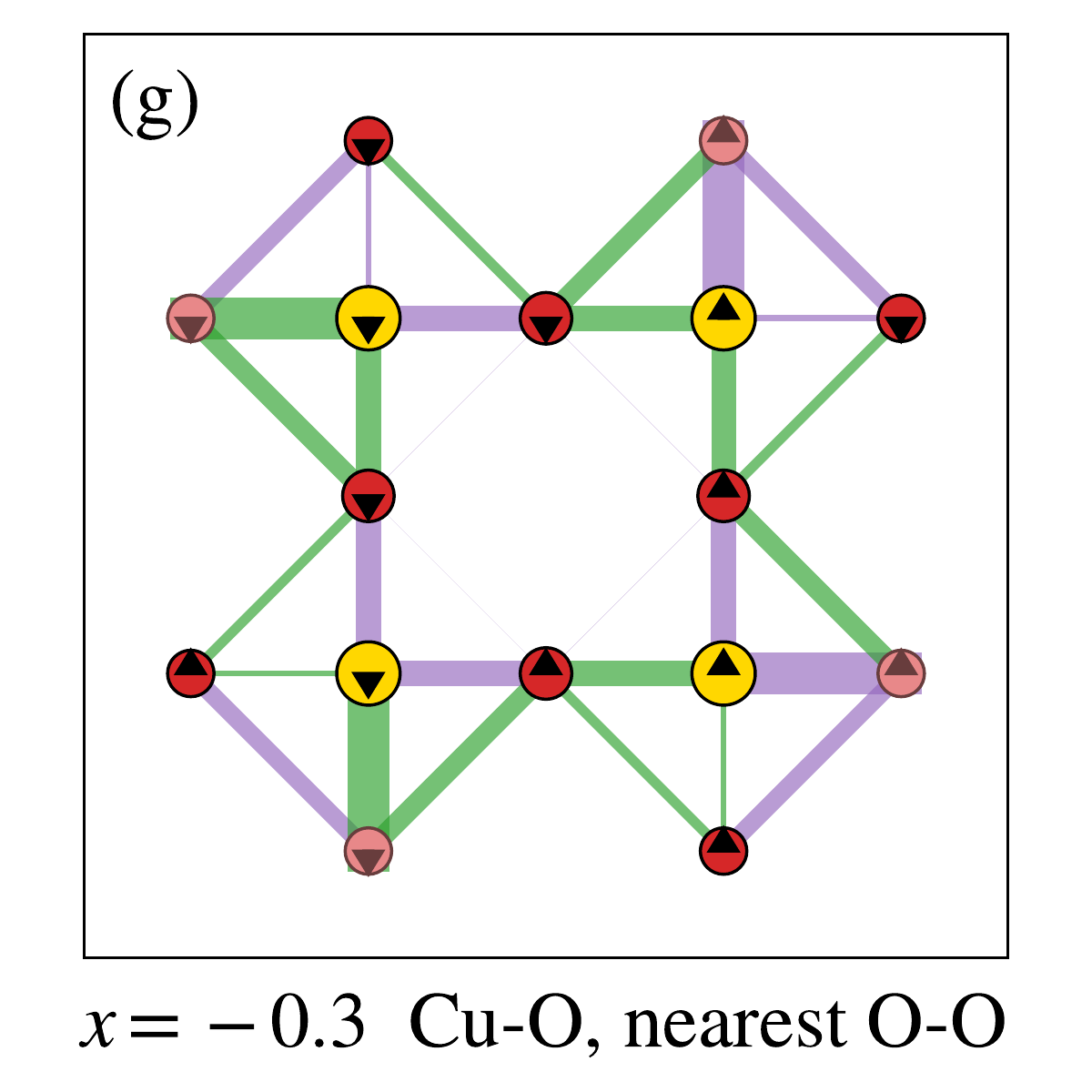}}
\caption{\label{fig: pattern}Charge, spin and pairing orders in the three-band Hubbard model. We use yellow and red circles for Cu and O respectively. The area of the circle reflects the corresponding local hole density, the length of the arrow denotes the magnitude of the local magnetic moment, the width of the lines is proportional to the pairing strength and different colors of the lines denote different coupling signs. The results are calculated based on the fully parametrized model at $x = 0.0$ [(a)], $x = 0.3$ doping [solution 1, (b) - (d)] and $x = -0.3$ doping [(e) - (g)]. (b) and (e) show the pairing strength between Cu and Cu; (c) and (f) show the pairing strength between the next nearest neighbor O; (d) and (g) illustrate the couplings of both the nearest Cu-O, and the nearest O-O. }
\end{figure*}

\subsubsection{Atomic scale orders in the full model} 
Beyond the bulk order parameters, the three-band model and the explicit inclusion of both copper and oxygen
atoms into the DMET impurity cluster allows for the possibility of studying the magnetic and superconducting order at the
atomic scale. The explicit charge, spin, and pairing orders are shown in Fig. \ref{fig: pattern}.
We only present representative results from the Hanke full model at $x = 0.0$, $x = 0.3$ (solution 1) and $x = -0.3$ doping, since  the  results from other parametrizations and
dopings are qualitatively similar.
(Further plots are presented in Figs. S3 - S8 of the
Supplemental Material~\cite{note-SI-3band}). 
Comparing Figs. \ref{fig:pattern a} and \ref{fig:pattern b}, we see that on doping the holes mainly occupy the oxygen sites and the hole density on copper only increases slightly. Combined with the fact that
doped electrons mainly reside on Cu [see Fig. \ref{fig:pattern e}, the hole density on Cu is reduced], this reflects the particle-hole asymmetry of the three-band model ~\cite{Lee06RMPhightc, White15}.
With respect to pairing order, we see $d_{x^2 - y^2}$-wave symmetry  clearly between neighboring Cu sites (i.e. it transforms according to the $B$-representation of the $C_4$ group and the sign of the pairing changes on rotating by $90^{\circ}$), see Figs. \ref{fig:pattern b} and \ref{fig:pattern e}.
The Cu-Cu pairing order is the largest pairing order between the atoms. From Figs. \ref{fig:pattern c} and \ref{fig:pattern f}, we also see  $d$-wave order between the \textit{next-nearest} O $p$ orbitals. Although the magnitude is slightly smaller than that of the Cu-Cu pairing, it still contributes almost $\approx 20\% - 40\%$
of the bulk $d$-wave order in  Eq. \eqref{eq: SC order param}. We note that the O-O pairing contribution is also asymmetric with respect to doping. In particular, its contribution can be as large as $\approx 40\%$ in the hole doped side but only 20-30\% in the electron doped region. 
Finally, we consider the pairing order between Cu-O and the nearest O-O atoms, see Figs. \ref{fig:pattern d} and \ref{fig:pattern g}. We see that the coupling between the nearest O-O atoms has $s$-wave symmetry but is quite weak, related to the incompatible orbital orientations.
On the other hand, we find the pairing between Cu-O to be relatively strong (in all parameter sets). The local symmetry of Cu-O coupling has $p_{x (y)}$-wave [or $d_{xy (yz)}$-wave] symmetry (the pattern transforms according to the $E$-representation of the $C_4$ group), which to our knowledge has
not previously been reported. We note that the superconducting phase pattern between Cu and O is similar to the orbital current-current correlation patterns in Ref. \cite{Thomale08-3band-ED-orb-current}, although the current-current correlations were reported to be extremely weak.
The pattern is also similar to the asymmetry reported as a hidden order in polarized elastic neutron diffraction experiments~\cite{Fauque06}.
Further investigation of these and other intriguing connections to intracell orders is left to future work.




\section{Conclusions}\label{sec: conclusion}

In summary, we have used density matrix embedding theory to characterize the ground-state phases of the three-band
Hubbard model. 
We have calculated the charge, local magnetic moments, projected energy bands and density of states of the undoped three-band model, which support
a charge-transfer insulating character at zero doping.  

We also studied
the doping dependence of the ground-state (phase diagram) of the model paying particular attention
to the local antiferromagnetic (AFM) and superconducting (SC) orders.
In a broad range of model parameters
we find a decrease in AFM order upon doping and a SC
dome. Unlike in the one-band picture, the models all predict a large region of coexistence of AFM + SC orders, with the
AFM order decreasing quite slowly.
Comparison to experimental data and earlier theoretical studies suggests that the minimal parametrized models overestimate
the AFM order and lead to poorer energy gaps, relative to the full parametrizations, which also include oxygen and oxygen-copper Coulomb repulsion, and
oxygen-oxygen hopping. The magnetic moment is particularly sensitive to the $\Delta_{pd}$ and $t_{pp}$ parameters, and  in the minimal model, the charge transfer gap $\Delta_{pd}$ should be renormalized downwards to better capture the experimental phase diagram. Within the full model, there are qualitative differences between the SC orders on the hole- and electron-doped side, with the electron-doped side supporting two different SC domes, one of which
appears more similar to the one observed in the one-band model, and the other like the SC dome observed on the hole-doped side.


The three-band model further allowed us to study order at the atomic scale. In the SC region, we observed strong $d$-wave pairing between Cu-Cu and the next-nearest O-O, weak extended $s$-wave coupling between the nearest O-O atoms,  and $p$- (or $d_{xz}, d_{yz}$)-like symmetry pairing between Cu-O. The intriguing symmetry of the latter order, similar to that seen in some experiments, illustrates the new physics
that emerges at atomic length-scales in the three-band model. Exploring such physics in more detail  will be the subject of future work.



\begin{acknowledgments}
  We thank Alec White, Nai-Chang Yeh, Patrick Lee, Tianyu Zhu and Yang Gao for helpful discussions. Z.-H.C. thanks Zhiao Yu for help on graphics. This work was supported by
  the US Department of Energy, Office of Science, via award No. 19390. G.K.C. is a Simons Investigator in Physics and is supported
  by the Simons Collaboration on the Many-Electron Problem.
\end{acknowledgments}

\appendix

\section{DMET Bath Construction}\label{app : DMET bath construction}

Given a Slater determinant lattice mean-field wavefunction $\Phi^{\rm HF}$, the DMET bath orbitals can be constructed in several equivalent ways, e.g.
via a singular value decomposition (SVD) of the MO coefficients \cite{Knizia12} or the environment-impurity part of the density matrix \cite{Zheng17}, or via eigenvalue decomposition of the projected overlap matrix \cite{Knizia13} or environment-environment block of the density matrix \cite{Wouters16}. All these methods define a set of bath orbitals, which has non-zero overlap with the impurity sites. In this work, we use SVD of the environment-impurity block of the density matrix to efficiently construct the bath in a periodic lattice system \cite{Zheng17, Cui20-dmet-solid}. By taking the first unit-  (or super-) cell as the impurity and the remaining cells as the environment, the whole density matrix of the lattice is divided into four blocks,
\begin{equation}
\gamma^{\Phi^{\rm HF}} = \begin{bmatrix}
\gamma^{\imp-\imp} &  \gamma^{\imp-\env} \\
\gamma^{\env-\imp} &  \gamma^{\env-\env} \\
\end{bmatrix} ,
\end{equation}
where $\gamma^{\imp-\imp}$ is the density matrix of the first cell, i.e. $\gamma (\vecR = \veczero)$; while $\gamma^{\env-\imp}$ is the coupling density matrix between the first cell and other cells, i.e. $\gamma (\vecR \neq \veczero)$. These two blocks can be easily computed in periodic systems by
a Fourier transform of the density matrix in $\veck$-space, 
\begin{equation}
\gamma^{\vecR}_{ij} = \frac{1}{N_{\veck}} \sum_{\veck} \ee^{\ii \veck \cdot \vecR} \gamma^{\veck}_{ij} ,
\end{equation}
where $N_{\veck}$ is the number of $\veck$-points in the first Brillouin zone. The bath can then be computed by SVD of $\gamma^{\vecR \neq \veczero}$ \cite{Cui20-dmet-solid}, 
\begin{equation}\label{eq: DMET bath use SVD}
\gamma^{\vecR \neq \veczero}_{ij} = \sum_{\tilde{i}} B^{\vecR \neq \veczero}_{i \tilde{i}} \Lambda_{\tilde{i} \tilde{i}} V^{\dg}_{\tilde{i} j} ,
\end{equation}
where $B$ yields the bath orbital coefficients and the singular values $\Lambda$ measure the entanglement between bath and impurity orbitals. We note that the $\gamma^{\imp-\env}$ and $\gamma^{\env-\env}$ blocks are not needed for bath construction and in fact, their computation and storage would be prohibitively expensive in a periodic calculation with many $\veck$ points. Therefore, using the SVD of $\gamma^{\vecR \neq \veczero}$  is more economical ($\mathcal{O} (N_{\veck} N^{3}_{\rm orb})$ cost) than diagonalizing the $\gamma^{\env-\env}$ block.


In addition to the cost, there are two other advantages of using the SVD. First, it is easy to discard the non-coupled bath orbitals, i.e.  bath orbitals with (almost) zero singular values. These bath orbitals are essentially core or virtual orbitals, which have little entanglement with the impurity, and they should be removed from the impurity problem for better numerical stability during the DMET self-consistency. Second,
when using  finite temperature smearing, the idempotency of the lattice mean-field density matrix is slightly broken. Rigorous treatment of finite-temperature requires a large number of bath orbitals than the number of impurity sites \cite{Sun20ftdmet}. However, the SVD still provides a good
first approximation of the finite-temperature bath, especially at low temperatures.

For the superconducting states $\Phi^{\rm HFB}$, the construction of the bath can still be carried out using SVD, but acting on the (env-imp block of)
of the generalized density matrix \cite{Zheng16},
\begin{equation}
\gamma^{\Phi^{\rm HFB}} = 
\begin{bmatrix}
\gamma^{\alpha \alpha} &  \kappa^{\alpha \beta} \\
\kappa^{\alpha \beta \dg}      &  I - \gamma^{\beta \beta T} \\
\end{bmatrix}
=
\begin{bmatrix}
\expval{a^{\dg}_{\alpha} a_{\alpha}} & \expval{a_{\beta} a_{\alpha}}\\
\expval{a^{\dg}_{\alpha} a^{\dg}_{\beta}}     &  I - \expval{a^{\dg}_{\beta} a_{\beta}}^T \\
\end{bmatrix} ,
\end{equation}
where the anomalous part of the density matrix $\kappa$ allows for a non-zero SC order parameter. Within the singlet pairing picture,
particles of one spin sector are allowed to couple with holes of the other spin. Therefore, the number of bath orbitals is effectively two times larger than in the normal state DMET.

\section{Analytic gradients of cost function Eq. \eqref{eq: cost function} at finite temperature}\label{app : gradient at finite T}

Once the gradients of Eq. \eqref{eq: cost function} are obtained, we can utilize
efficient gradient-based numerical methods, such as CG or the
Broyden-Fletcher-Goldfarb-Shanno (BFGS) algorithm, to optimize the correlation
potential.
By differentiating Eq. \eqref{eq: cost function} with respect to $u_{ij}$ we have, 
\begin{equation}\label{eq : cost function derivative}
\pdv{w}{u_{ij}} = 2 \sum_{kl} \qty(\gamma^{\rm mf} - \gamma^{\rm corr})_{kl} \pdv{\gamma^{\rm mf}_{kl}}{u_{ij}} ,  
\end{equation}
and thus the key task in Eq. \eqref{eq : cost function derivative} is to evaluate the response of the mean-field density matrix with respect to a perturbation, $\partial \gamma^{\rm mf}_{kl} / \partial {u_{ij}} $. The response at zero temperature can be written in terms of orbital coefficients and energies
(see e.g. Refs. ~\cite{Wouters16, Zheng18}) using first order perturbation theory, 
\begin{equation}\label{eq : derivative 0 K}
\pdv{\gamma^{\rm mf}_{kl}}{u_{ij}} = \sum_{p}^{\rm occ} \sum_{q}^{\rm virt}\frac{C_{kp}C^{*}_{lq}C_{iq}C^{*}_{jp} - C_{kq}C^{*}_{lp}C^{*}_{iq}C_{jp}}{\epsilon_p - \epsilon_q},
\end{equation}
where we have assumed the system is \textit{gapped}. However, when the system becomes (nearly) gapless, this expression diverges.
In such cases, the divergent gradient  causes the optimization to fail, and this
is a source of many convergence difficulties in DMET.

One way to ameliorate this issue is to introduce a finite temperature smearing, similar to what is used in mean-field calculations of metals. With an inverse temperature $\beta$ and a perturbation $\delta u$, the Fermi-Dirac density matrix is defined as,
\begin{equation}\label{eq : dm at finite T}
\gamma_{kl} = \qty[1 + {\rm e}^{\beta \qty({h} - \mu + \delta u)}]^{-1}_{kl},
\end{equation}
where $\mu$ is the Fermi level for the (quasi-)particles.
The response of $\gamma$ with respect to the correlation potential $u$ then involves two terms,
\begin{equation}\label{eq: drho dv two terms}
\begin{split}
\dv{\gamma_{kl}\qty[u, \mu(u)]}{u_{ij}} 
& = \left.\pdv{\gamma_{kl}}{u_{ij}} \right|_{\mu} + \pdv{\gamma_{kl}}{\mu} \pdv{\mu}{u_{ij}} ,
\end{split}
\end{equation}
where the first term is the direct response of the density at a fixed Fermi level,
while the second term reflects the contribution of the implicit change in the Fermi
level due to the change in potential. The final expression for the first term
in Eq. \eqref{eq: drho dv two terms} is,
\begin{equation} \label{eq : derivative final}
\pdv{\gamma_{kl}}{u_{ij}} = \sum_{pq} C_{kp} C_{ip}^{\ast} K_{pq} C_{jq} C_{lq}^{\ast} ,
\end{equation}
where
\begin{equation}\label{eq: Kpq}
K_{pq} \equiv n_p \qty(1 - n_q) \frac{1 - {\rm e}^{\beta(\varepsilon_p - \varepsilon_q)}}
	{\varepsilon_p - \varepsilon_q}.
\end{equation}
It is easy to check that $K_{pq}$ is always finite when $\varepsilon_{p} =
\varepsilon_{q}$. One can also let $\beta$ go to infinity and choose $p$ / $q$
to label occupied / virtual orbitals; the gradient then gives the correct zero
temperature limit in Eq. \eqref{eq : derivative 0 K} (up to a symmetrization).

The final expressions for the second term in Eq. \eqref{eq: drho dv two terms} are,
\begin{equation}\label{eq: mu contribution 2}
\begin{split}
\pdv{\gamma_{kl}}{\mu} &= \sum_{p} \beta C_{kp} n_p (1-n_p) C^{*}_{lp}, \\
\pdv{\mu}{u_{ij}} &= \qty[\sum_{p} n_p (1 - n_p) C^{*}_{ip} C_{jp}] / \qty[\sum_{p} n_p (1 - n_p)].
\end{split}
\end{equation}
Usually this contribution is very small at low temperatures, compared to the direct
response in Eq. \eqref{eq : derivative final}. However, this contribution will be important in a real finite temperature simulation, e.g. in Ref. ~\cite{Sun20ftdmet}.

We summarize the derivation of Eqs. \eqref{eq : derivative final} - \eqref{eq: mu
contribution 2} in the Supplemental Material~\cite{note-SI-3band}.

\section{Numerical convergence}\label{app: convergence}



Here we assess the accuracy and convergence of the DMET procedure in the three-band model calculations.
The error in the DMET calculations arises from three possible sources: 
(a) DMET self-consistency error (from incomplete convergence), 
(b) DMRG solver error due to the finite bond dimension, and 
(c) error from the finite size of the impurity. 
The finite size error (c) can, in principle, be eliminated by increasing the
cluster size and extrapolating to the thermodynamic limit (TDL), as performed in
the one-band Hubbard model case ~\cite{Zheng16}.
In this work, we use a fixed $2 \times 2$ cluster size due to the increased computational cost of the three-band model, and thus we cannot assess the finite-size error, except via some comparisons to the $2\times 2$ cluster error in the one-band model.
However, the error due to (a) and (b) can be directly estimated in our framework, which we now discuss.

\begin{figure}[!htb]
\includegraphics[width=85mm, clip]{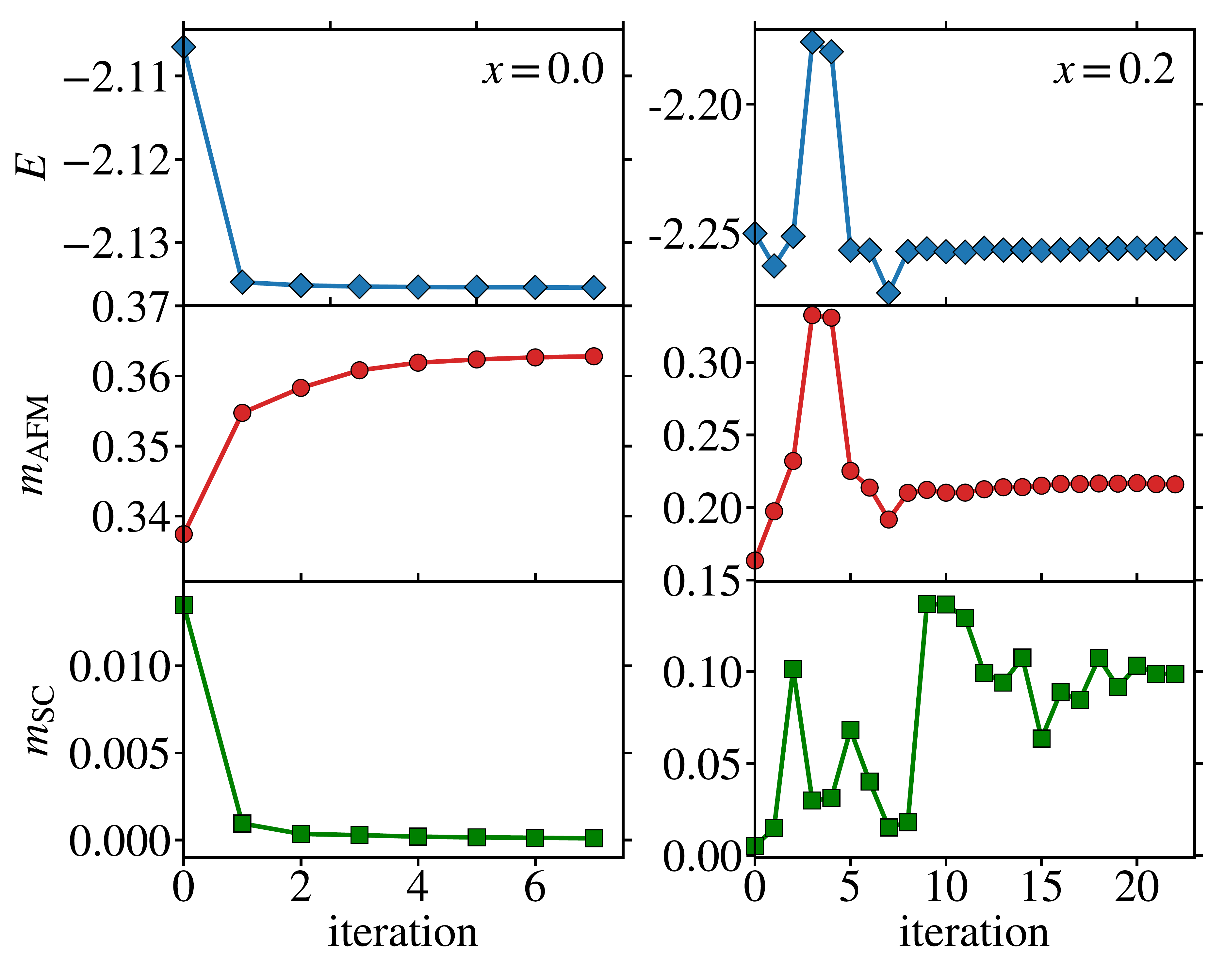}
\caption{\label{fig:conv-vs-iter} DMET energy (in units of $t_{pd}$) and order parameters 
of the Hybertsen minimal parametrized three-band model, with respect to the number of iterations, 
at doping $x = 0.0$ (left) and $x = 0.2$ (right).}
\end{figure}
Fig. \ref{fig:conv-vs-iter} shows the overall convergence of DMET with respect to the number of DMET self-consistent iterations.
We observe qualitatively different convergence in the normal and superconducting parts
of the DMET phase diagram. To illustrate this, we plot
the DMET energy, AFM, and ($d$-wave) SC order parameter for the Hybertsen model at different dopings $x$.
(These order parameters are defined precisely in Sec.~\ref{subsec: computational details}). We first discuss the undoped system.
Here we see that the DMET cycle converges smoothly within 7 iterations. For the DMET energy, a single DMET step is enough to converge
to  $\approx 10^{-4}$, demonstrating the utility of single-shot DMET calculations in normal (and especially non-magnetic) states.
The order parameters (density matrices)
are more strongly affected by self-consistency. We find that the AFM order increases during the iterations, while the SC order is suppressed,
giving a pure antiferromagnetic state at convergence. We next consider $x = 0.2$ doping. Here, the
self-consistency cycle converges more slowly, requiring about 20 DMET iterations to reach convergence. 
The total energy as well as AFM order converges at around the \nth{10}
iteration, while the SC order oscillates until the \nth{20} iteration. This in
part reflects the influence of the initial guess: the AFM guess [$v^{\sigma}$ in
Eq. \eqref{eq: u potential}] is quite close to the converged potential, while
the SC guess [$\Delta^{\alpha\beta}$ in Eq. \eqref{eq: u potential}] is initialized randomly and thus needs more iterations to converge. 
If we were to restrict
the DMET optimization to only pairing potentials with $d$-wave symmetry (as is commonly done in most cluster
DMFT~\cite{Senechal08} or VCA calculations~\cite{Hanke10}), the convergence would be much faster. 
However, the more general form of the correlation potential in DMET allows for the possibility of other pairing channels 
and orders to emerge.
  The remaining DMET self-consistency error can be estimated from the
  difference between the expectation values (e.g. DMET energy) of the last two
  iterations ~\cite{Zheng16}, e.g. $ \delta E = \frac{1}{2} \qty| E(n-1) - E(n) | $. 
  Consistent with our chosen convergence criterion,
  the typical size of the DMET self-consistency error in the undoped region is less than $10^{-5}$ (for both the energy and order parameters),
  and less than $10^{-4}$ (for the energy) and $\sim 10^{-3}$ (for the order parameters) in the doped region.
   
\begin{figure}[!htb]
\includegraphics[width=85mm, clip]{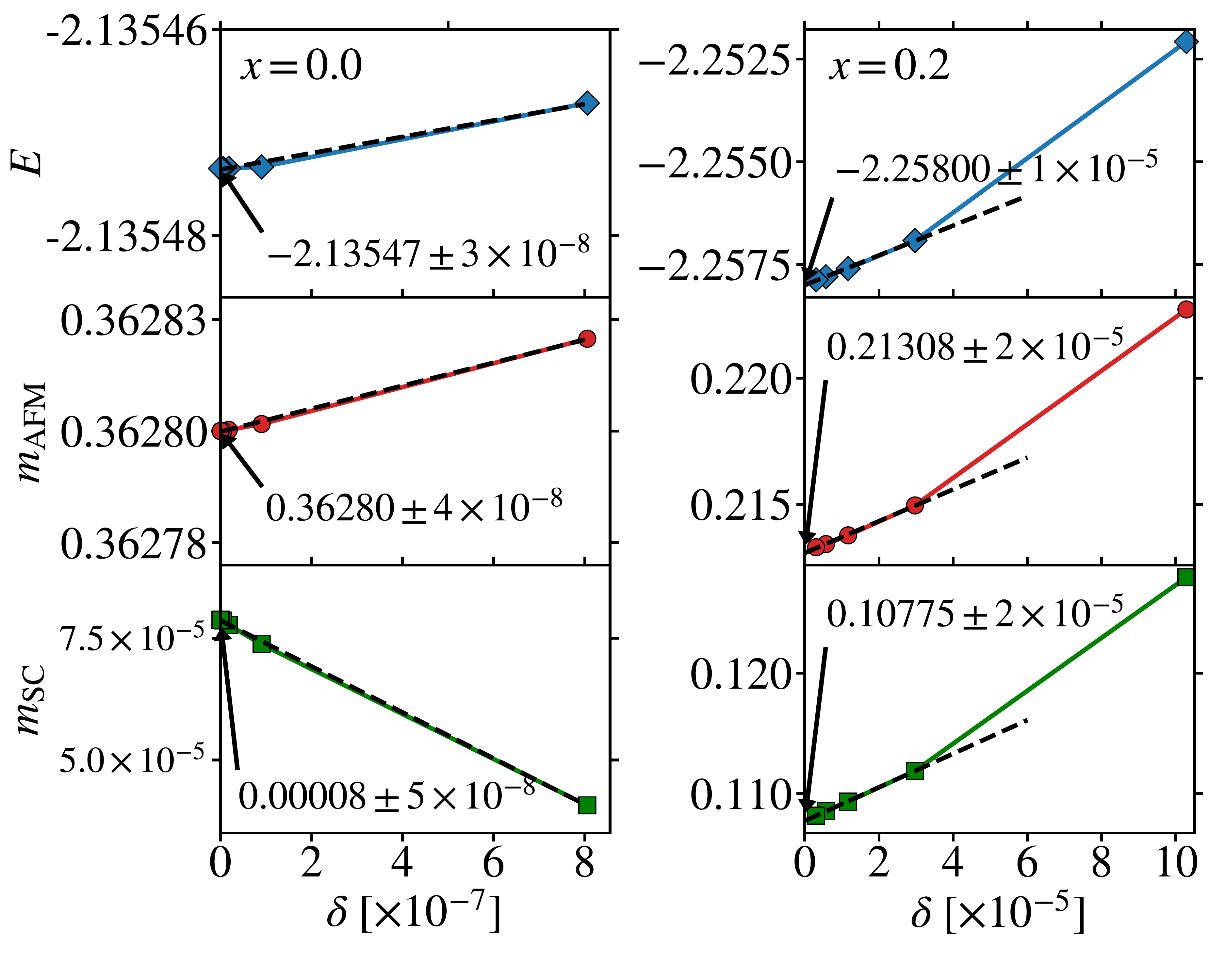}
\caption{\label{fig:conv-vs-M} DMET energy (in units of $t_{pd}$) and order
parameters of the Hybertsen minimal parametrized three-band model, with respect to the discarded weight $\delta$ of the DMRG solver, 
at doping $x = 0.0$ (left) and $x = 0.2$ (right). The values are linearly extrapolated to the limit where $\delta = 0.0$ (dashed line). The error shown is the standard deviation of linear regression.}
\end{figure}
The error from the DMRG solver can be estimated using standard techniques based on the 
discarded weight in the DMRG calculation~\cite{Legeza96, Chan02, White07} and can be further reduced by extrapolation.
  The error in the impurity observables (used to evaluate the DMET
energy and order parameters) is linear in the (sufficiently small) discarded weight $\delta$ and hence can be extrapolated to the exact result ($\delta=0$)~\cite{White07}.
The convergence with bond dimension  $M$ for fixed correlation potential $u$ is shown in Fig. \ref{fig:conv-vs-M}.
We find that the discarded weight in the normal state (undoped model) is extremely
small and usually less than $10^{-8}$, thus extrapolation is unnecessary. In fact, calculations can be
carried out using a bond
dimension as small as $M = 100$ without any significant error. On the other hand, when the system becomes superconducting,
the discarded weight also increases, e.g. to $3 \times 10^{-5}$ at $M = 800$, indicating
that the system is more entangled.
In such situations, extrapolation has a significant effect on the DMET expectation values.
Compared to the extrapolated values, at $M = 800$ the error in the energy (per site) and
order parameters is about $10^{-3}$. 


In summary, from the above analysis, we find that the DMET calculations can be smoothly converged, with minimal error  from either the self-consistency or from the solver.


\bibliography{refs-3band}
\end{document}